\documentclass[journal]{vgtc}                

\ifpdf
  \pdfoutput=1\relax                   
  \usepackage{graphicx}                
  \DeclareGraphicsExtensions{.pdf,.jpg,.jpeg} 
\else
  \usepackage{graphicx}                
  \DeclareGraphicsExtensions{.eps}     
\fi%

\graphicspath{{figures/}{pictures/}{images/}{./}} 

\usepackage{microtype}                 
\PassOptionsToPackage{warn}{textcomp}  
\usepackage{textcomp}                  
\usepackage{times}                     
\usepackage{cite}                      
\usepackage{tabu}                      
\usepackage{booktabs}                  

\usepackage{graphicx}
\usepackage{color,colortbl}					
\usepackage{afterpage}
\usepackage{subfigure}
\usepackage{slantsc}
\usepackage{amsmath}
\usepackage{algorithm}
\usepackage{algpseudocode}
\usepackage{mleftright}
\usepackage{amssymb}
\usepackage{tabularx}
\usepackage{hhline}
\usepackage{tikz}
\usepackage{stackengine}
\def\mathbi#1{\textbf{\em #1}}
 
\newcommand{\eps}{\mbox{$\epsilon$}}

\def\x{\mathbi{x}}

\let\vec=\mathbi%
\let\mat=\mathbf%
\let\set= \mathcal%

\definecolor{RED}{rgb}{1,0,0}

\newcommand{\specialcell}[2][c]{%
  \begin{tabular}[#1]{@{}c@{}}#2\end{tabular}}

\onlineid{0}

\vgtccategory{Research}
\vgtcpapertype{please specify}

\title{TTHRESH: Tensor Compression for Multidimensional Visual Data}

\author{Rafael~Ballester-Ripoll,~\textit{Member,~IEEE,}
        Peter~Lindstrom,~\textit{Senior~Member,~IEEE},\\
        and~Renato~Pajarola,~\textit{Senior~Member,~IEEE}}
\authorfooter{
\item
 R. Ballester-Ripoll and R. Pajarola are with the Department
 of Informatics, University of Z\"{u}rich, Switzerland. E-mails: rballester@ifi.uzh.ch and pajarola@ifi.uzh.ch.
 
 \item Peter Lindstrom is with the Center for Applied Scientific Computing, Lawrence Livermore National Laboratory, USA. E-mail: pl@llnl.gov.
}

\shortauthortitle{Ballester-Ripoll \MakeLowercase{\textit{et al.}}: TTHRESH: Tensor Compression for Three-dimensional Visual Data}

\abstract{Memory and network bandwidth are decisive bottlenecks when handling high-resolution multidimensional data sets in visualization applications, and they increasingly demand suitable data compression strategies. We introduce a novel lossy compression algorithm for multidimensional data over regular grids. It leverages the higher-order singular value decomposition (HOSVD), a generalization of the SVD to three dimensions and higher, together with bit-plane, run-length and arithmetic coding to compress the HOSVD transform coefficients. Our scheme degrades the data particularly smoothly and achieves lower mean squared error than other state-of-the-art algorithms at low-to-medium bit rates, as it is required in data archiving and management for visualization purposes. Further advantages of the proposed algorithm include very fine bit rate selection granularity and the ability to manipulate data at very small cost in the compression domain, for example to reconstruct filtered and/or subsampled versions of all (or selected parts) of the data set.} 


\CCScatlist{ 
 \CCScat{K.6.1}{Management of Computing and Information Systems}%
{Project and People Management}{Life Cycle};
 \CCScat{K.7.m}{The Computing Profession}{Miscellaneous}{Ethics}
}

\teaser{
  \centering
  \subfigure[Original (512MB)]{\includegraphics[width=0.28\linewidth]{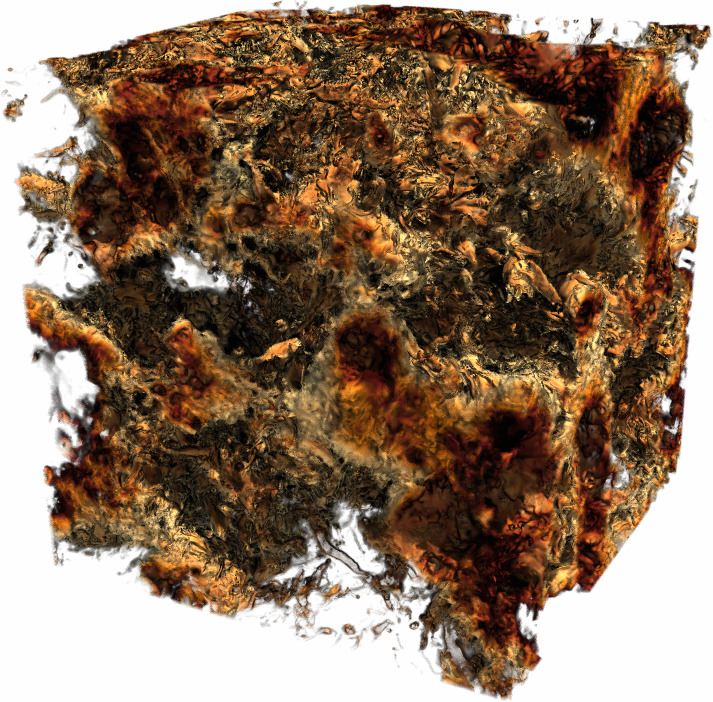}} \hfil
  \subfigure[10:1 compression (51.2MB)]{\includegraphics[width=0.28\linewidth]{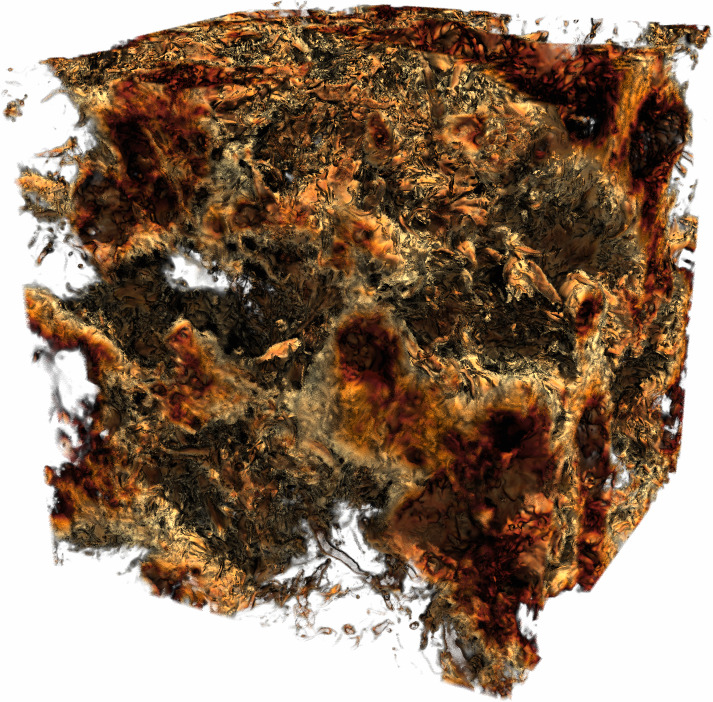}} \hfil
  \subfigure[300:1 compression (1.71MB)]{\includegraphics[width=0.28\linewidth]{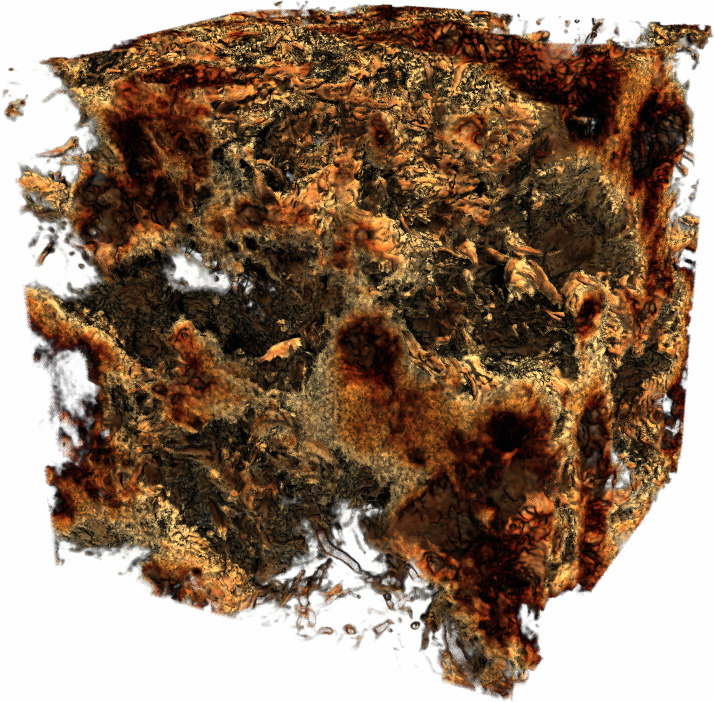}}
  \caption{(a) a $512^3$ isotropic turbulence volume~\cite{jhtd}; (b) visually identical compression result; (c) result after extreme compression.}
	\label{fig:teaser}
}

\vgtcinsertpkg

\begin{document}




\firstsection{Introduction}\label{sec:introduction}

\maketitle


%
%
%
%


Most scientific and visual computing applications face heavy computational and data management challenges when handling large and/or complex data sets over Cartesian grids. Limitations in memory resources or available transmission throughput make it crucial to reduce and compress such data sets in an efficient manner. Lossy compression is often the prescribed strategy, since many applications admit a certain error (especially for higher bit depths and floating-point precision). If the compressed data set is to be used for subsequent computational analysis and/or to be fed as the initial state of a simulation routine, only small errors are typically tolerated. Conversely, if visualization and user exploration are to follow decompression, then higher error rates are acceptable; the method developed in this paper is mainly geared towards this case. Depending on the specific application, certain additional properties are sometimes desired. These may include fast support for random-access decompression, fine compression rate granularity, asymmetry (faster decompression than compression), bounded error, support for arbitrary dimensionality, ease of parallelization, topological robustness, etc. These aspects make multidimensional compression a broad and challenging problem for which, unsurprisingly, no catch-all solution exists.

In this context, \emph{tensor decompositions} and in particular the \emph{Tucker model} are promising mathematical tools for higher-order compression and dimensionality reduction in the fields of graphics and visualization. 3D scalar field compression at the Tucker transform coefficients level was recently investigated~\cite{BP:15}, and it was concluded that coefficient thresholding outperforms earlier rank truncation-based approaches in terms of quality vs.\ compression ratio. This has motivated us to develop and introduce \textsc{tthresh}, a novel lossy compressor based on the Tucker decomposition. It is the first of its kind that supports arbitrary target accuracy via bit-plane coding. Previous related approaches fixed a number of quantization bits per transform coefficient, and sometimes even the transform basis size (the \emph{tensor ranks}). Instead, our method drastically improves the compression ratio-accuracy trade-off curve by greedily compressing bit planes of progressively less importance. We also extend our encoding scheme to compress the factor matrices. The importance of this is unique to the HOSVD transform, which needs to store its learned bases as opposed to fixed-basis methods, yet never optimized by earlier works.

We also benchmark \textsc{tthresh} against other state-of-the-art compressors that are not based on the HOSVD. While the ratios we achieve at low error tolerances are comparable to those, we significantly outperform them on the higher error ranges on which visualization tasks usually rely.

We have released an open-source C++ implementation of our algorithm\footnote{Available (LGPL-3.0) at \url{https://github.com/rballester/tthresh}.}. It is primarily intended as a standalone command-line utility, although its main functions are also usable in a header-only library fashion.

\section{Related Work}

\subsection{3D Compression Algorithms}

A number of lossy compression algorithms for scientific volume data sets have been proposed in the recent literature. For instance, \textsc{isabela}~\cite{LSEKLRS:11} focuses on spatio-temporal data with ample high-frequency components; it proceeds by sorting elements into a monotonic curve which is then fitted using B-splines. A more recent example of linearization strategy is \textsc{sz}~\cite{DC:16}, which either predicts each coefficient using low-degree polynomials on preceding coefficients, or truncates it in its IEEE 754 binary representation. Some methods prioritize preserving specific properties of the data set, for example bounded error using topological features~\cite{SPCT:18} or over connected and coherent regions (e.g. \textsc{sq}~\cite{IKK:12}). Vector quantization~\cite{GIM:12, GG:16} requires heuristics or greedy algorithms during compression, but is fast to decompress and thus suitable for compression-domain direct volume rendering; see also the survey~\cite{BGIMMPS:13}. In particular,~\cite{GIM:12} was defined within an octree multiresolution hierarchy for fast ray-casting in an interactive volume visualization application.

A popular and long-standing family of compression methods are the ones that exploit linear transforms, including well-known decompositions such as the Fourier and discrete cosine transforms~\cite{YL:95} and, since the 1990s, wavelets~\cite{Muraki:93,GS:01,NS:02,GWGS:02,WQ:05}. They capitalize on carefully designed transform bases that aim to sparsify real-world signals as much as possible. \textsc{vapor}~\cite{CMNR:07}, for example, uses a flexible wavelet-based compression layer integrated into an interactive volume and flow exploration tool. \textsc{zfp}~\cite{Lindstrom:14} is a floating-point compressor that uses custom transform matrices and emphasizes fast random access and low error for, among other applications, storing snapshots and intermediate steps in a computational analysis/processing pipeline. \textsc{zfp} offers a transparent interface for compressed C/C++ arrays and operates via fixed-rate encoding, although a variable-rate variant is also supported.

\subsection{Compressors Based on Tensor Decomposition}

Several transform-based compression algorithms have been recently proposed that use data-dependent bases (the so-called \emph{factor matrices}) instead of predefined ones. This is precisely the idea behind principal component analysis (PCA) as well as the Tucker decomposition. The Tucker model seeks to improve transform-domain sparsity at the expense of having to store its learned bases, which tends to be comparatively small for a three or more dimensions. Some of the earliest Tucker-based compression approaches for visual data include \cite{WA:04}, \cite{WXY:07b} and \cite{WXCLWY:08}. Progressive tensor rank reduction (the so-called \emph{truncation}; see later sections) has been shown to reveal features and structural details at different scales in volume data~\cite{SZP:10}. Further recent efforts in the context of tensor compression include~\cite{SIMAEZGGP:11,BGIMMPS:13,SMP:13,BSP:15,BP:15} for interactive volume rendering and visualization,~\cite{WLHR:12} for 3D displays,~\cite{BP:18} for integral histograms of images and volumes, and \cite{RK:09,Tsai:09,TS:12,Tsai:15} for reflectance fields, among others. The large-scale renderer \textsc{tamresh}~\cite{SMP:13} resembles block-transform coding in that the input volume is partitioned in small multiresolution cubic bricks; each brick is then compressed as a separate HOSVD core. Recently, Tucker core hard thresholding combined with factor matrix quantization was shown~\cite{BP:15} to yield better compression rate than slice-wise truncating the core. These points have motivated the compressor proposed here.

\section{Tucker/HOSVD Decomposition} \label{sec:tucker_compression}

Throughout this paper, tensors refer to multiarrays of dimension $N \ge 1$. We write vectors (tensors of dimension 1) in bold lowercase as in $\vec{x} = (x_1, \dots, x_N)$, matrices (tensors of dimension 2) in bold capitals such as $\mat{U}$, and general tensors as well as sets in calligraphic letters such as $\set{T}$. We generally use the notation and definitions from~\cite{LMV:00b}; in particular, rows and columns in matrices generalize to tensors as \emph{fibers}. The \emph{$n$-th mode unfolding} of a tensor $\set{T}$ arranges all $n$-mode fibers next to each other as columns of a wide matrix and is denoted as $\mat{T}_{(n)}$. The \emph{tensor-times-matrix product} (TTM) contracts a tensor's $n$-mode fibers along a matrix's rows and is denoted as $\set{T} \times_n \mat{U}$. We access tensors using bracket notation, so for instance $\mat{U}[1, 1]$ is the top left element of a matrix $\mat{U}$. We refer the reader to Kolda and Bader's survey~\cite{KB:09} for more extensive details on basic tensor manipulation.

\subsection{The Tucker Model} \label{sec:tucker_model}

\begin{figure*}[ht]
\centering
\subfigure[Core truncation]{\includegraphics[width=0.65\columnwidth]{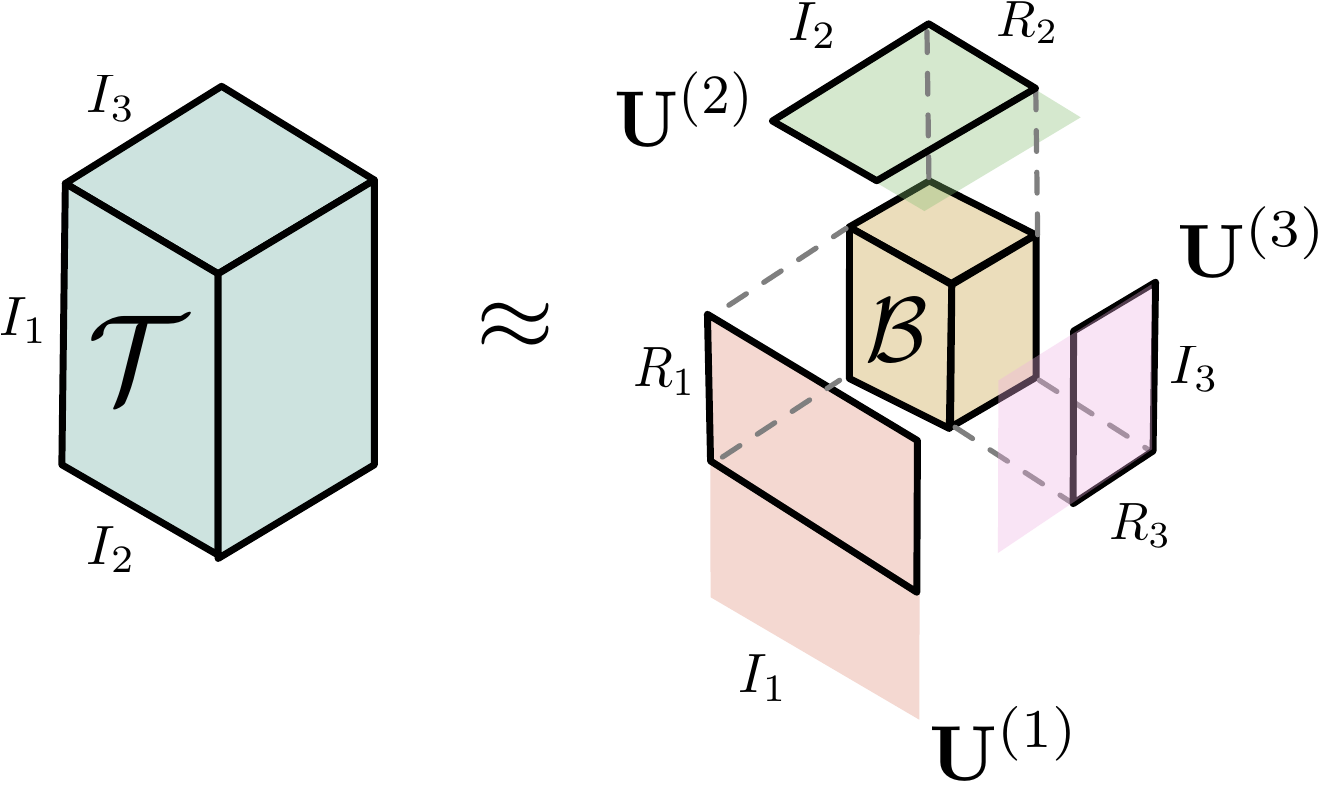}} \hfil
\subfigure[Full core, used in this paper for subsequent processing and quantization]{\includegraphics[width=0.65\columnwidth]{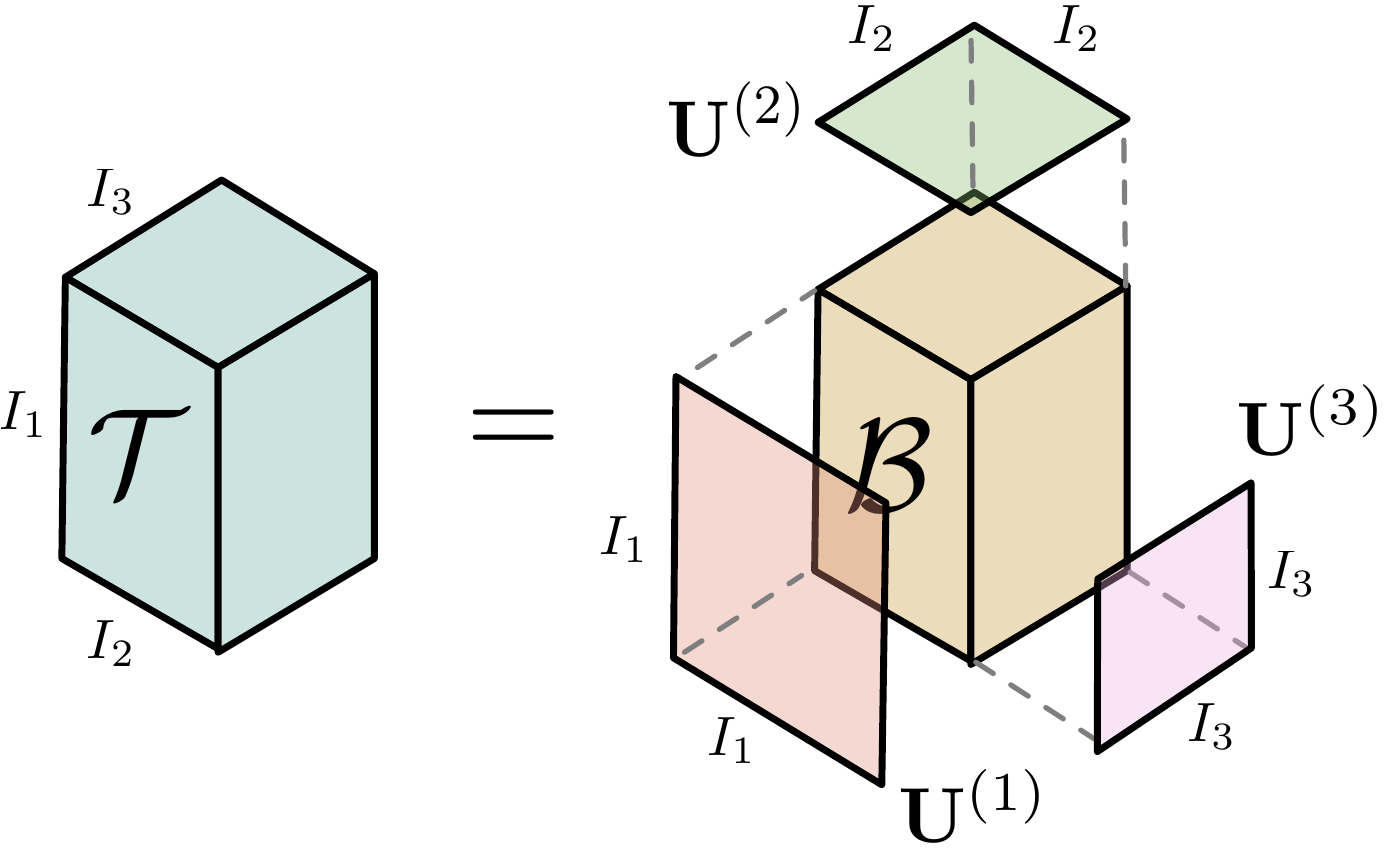}}
\caption{Left: the Tucker rank truncation approach for 3D compression used in e.g.~\cite{WXCLWY:08}, \cite{SIMAEZGGP:11}, \cite{SMP:13} and \cite{BSP:15}. Right: the full core approach first considered in~\cite{BP:15} and here extended into a full-fledged compressor with adaptive thresholding and bit-plane coding.}
\label{fig:tucker}
\end{figure*}

The \emph{full} Tucker decomposition~\cite{Tucker:66, LMV:00b} writes any entry $\set{T}[x_1, \dots, x_N]$ of a 3D tensor $\set{T}$ exactly as:
\begin{equation} \label{eq:hosvd2}
\sum_{r_1, \dots, r_N=1}^{I_1, \dots, I_N} \set{B}[r_1, \dots, r_N] \cdot \mat{U}^{(1)}[x_1, r_1] \cdots \mat{U}^{(N)}[x_N, r_N]
\end{equation}
or, in the more compact TTM notation,
\begin{equation} \label{eq:tucker}
\set{T} = \set{B} \times_1 \mat{U}^{(1)} \times_2 \dots \times_N \mat{U}^{(N)}
\end{equation}
where each $\mat{U}^{(n)}$ is a non-singular matrix of size $I_n \times I_n$ and $\set{B}$ is a \emph{core} tensor of coefficients with the same size as $\set{T}$. See Fig.~\ref{fig:tucker}(b) for an illustration of the full Tucker decomposition. The matrices $\mat{U}^{(n)}$ are called \emph{Tucker factors} (or \emph{factor matrices}) and define a two-way transformation between $\set{T}$ and its core $\set{B}$, whereby Eq.~\ref{eq:tucker} is inverted as
\begin{equation}
\set{B} = \set{T} \times_1 {\mat{U}^{(1)}}^{-1} \times_2 \cdots \times_N {\mat{U}^{(N)}}^{-1}.
\end{equation}

The higher-order singular value decomposition (HOSVD)~\cite{LMV:00b, KB:09} is an efficient procedure to construct orthogonal Tucker factors (i.e., whose columns are orthogonal unit vectors) by setting each $\mat{U}^{(n)}$ as the left singular vectors of the $n$-th mode unfolding matrix $\mat{T}_{(n)}$. In other words, the HOSVD sets each $n$-th factor as the uncentered PCA transformation matrix of the set of all fibers from $\set{T}$, taken along the $n$-th mode. The Tucker model is flexible and readily applicable to any shape and dimensionality, and the HOSVD decomposition always exists.

Since for three and more dimensions the core holds far more coefficients than the factors, it is also the decomposition part where most data reduction can be achieved and, consequently, the main source of error. Fortunately, we can determine and bound the $l^2$ error (i.e. sum of squared errors, or SSE for short) that is due to the core by just looking at its coefficients. Factor orthogonality implies that $\|\set{T}\| = \|\set{B}\|$. Furthermore, any perturbation in the core propagates directly to the reconstruction: $\|\widetilde{\set{B}}-\set{B}\| = \|\widetilde{\set{T}}-\set{T}\|$; see e.g.~\cite{KB:09, LMV:00b}. This property will be crucial for our compression strategy.

\subsection{Sparsifying Properties}

Tucker-based compression algorithms exploit the fact that the HOSVD transform coefficients generated in $\set{B}$ tend to be quasi-sparse for typical real-world or simulated multidimensional signals.
In addition, many transformations do not significantly affect the HOSVD. For example, if one permutes some slices of $\set{T}$ along one or more dimensions, its HOSVD will produce the same core $\set{B}$ and factors (with their corresponding rows permuted). Other possible transformations that can be encoded on a HOSVD-compressed data set without essentially affecting $\set{B}$ include spatially moving or stretching the data, padding it with zeros, upsampling it with multilinear interpolation, scaling by a constant (this will scale $\set{B}$), etc. Many usual data reduction approaches are guaranteed to actually improve HOSVD core sparsity, including downsampling, box-filtered decimation, convolving with any band-limited kernel, etc. For instance, a volume whose $k$ last wavelet levels are zero can be represented using a Tucker core with $8^k$ times fewer non-zero coefficients than otherwise needed.

The HOSVD decomposition decorrelates the data at all spatial scales, but does so without explicit space partitioning, i.e. avoiding tree-like structures or predefined multiresolution filter banks. The task of capturing correlation at multiple scales is thus undertaken by the different factor matrix columns. Nonetheless, the question of how to organize the coefficients in $\set{B}$ for effective compression is unclear \emph{a priori}. 

\subsection{Core Truncation and Its Limitations}

Conveniently, the HOSVD produces core (hyper-)slices that are non-increasing in norm. Let us consider the norm of each $k$-th slice of the Tucker core along the $n$-th dimension:
\begin{equation}
\sigma_k^{(n)} := \|\set{T}[\overbrace{:, \dots, :}^{n-1}, k, \overbrace{:, \dots, :}^{N-n}]\|.
\end{equation}

These norms have been proposed as \emph{generalized singular values}, and they satisfy~\cite{LMV:00b}:
\begin{equation} \label{eq:non_decreasing}
\sigma_1^{(n)} \ge \sigma_2^{(n)} \ge ... \ge \sigma_{I_n}^{(n)} \ge 0.
\end{equation}

Furthermore, from the factor matrix orthogonality it follows that the mean squared error (MSE) induced by zeroing-out a core coefficient is proportional to its squared magnitude. These properties have been exploited in the past as the basis of several truncation-based HOSVD compression schemes~\cite{WXCLWY:08}, \cite{SIMAEZGGP:11}, \cite{SMP:13}, \cite{BSP:15}, whereby the least important trailing factor columns in each $\mat{U}^{(n)}$ and the corresponding core slices in $\cal{B}$ along each dimension are discarded to produce a compressed approximation. By doing so, only $1 \le R_n < I_n$ factor columns and core slices remain for each mode $n$ (see Fig.~\ref{fig:tucker}(a)). The quantities $R_1, \dots, R_N$ are known as \emph{truncated Tucker ranks} and were used in those works for variable-detail compression and progressive reconstruction.

\begin{figure}[ht]
\centering
\subfigure{\includegraphics[width=0.4\columnwidth]{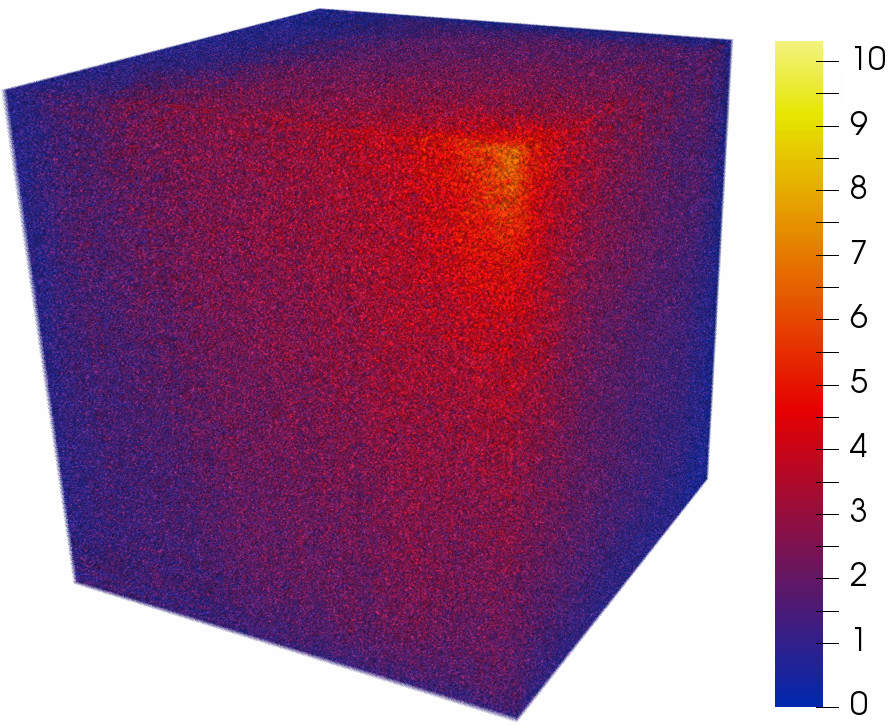}}
\caption{HOSVD core $\set{B}$ of size $256^3$, obtained from the Foot data set. For visualization we scale all values $x \mapsto \ln (1+x)$, then apply the colormap shown on the right. Note the \emph{hot corner} phenomenon.}
\label{fig:core_rendering}
\end{figure}

There are, however, two notable aspects that have not been pursued satisfactorily by these previous approaches. First, although the slice truncation idea is sound as motivated by Eq.~\ref{eq:non_decreasing}, its granularity is very coarse. Elimination strategies on a coefficient-by-coefficient basis (rather than slice-by-slice) have the potential to significantly improve compression quality. Second, and regardless of the coefficient elimination method chosen, how to encode the surviving coefficients remains an open issue as well. Based on the roughly exponential growth of those coefficients, Suter et al.~\cite{SIMAEZGGP:11} proposed a fixed-bit logarithmic quantization scheme: 1 bit for the coefficient sign and 8 or 16 for the logarithm of its absolute value. The authors realized the extreme importance of the first element $\set{B}[1, 1, 1]$ (the so-called \emph{hot corner}, as shown in Fig.~\ref{fig:core_rendering}); it often captures most of the signal's energy and $\|\set{B}[1, 1, 1]\| \approx \|\set{T}\|$. Hence this value was saved separately at 64-bit floating-point precision. This strategy was later replicated in other works~\cite{SMP:13,BP:15}. Nevertheless, a truly adaptive compression approach for the full-length HOSVD core has not been explored as of yet. The strategy we propose builds on the thresholding-oriented analysis of~\cite{BP:15} in that we compress coefficients, one bit plane $p$ at a time, up to a certain plane $63 \ge P \ge 0$. In particular, elements whose absolute value is below $2^P$ are thresholded away. We give the full details in the following section; see also Fig.~\ref{fig:tucker}(b).

\section{Proposed Algorithm}

Let $\set{T}$ be an input tensor and $\widetilde{\set{T}}$ the result after compression and decompression. Our pipeline accepts one main compression parameter, namely the error target, which can be specified in one of three ways:

\begin{itemize}
	\item Relative error (\texttt{-e} flag; sometimes known as \emph{normalized root mean square error}):
	$$\eps(\set{T}, \widetilde{\set{T}}) := \|\set{T} - \widetilde{\set{T}}\| / \|\set{T}\|,$$
	where $\|\cdot\|$ denotes the Frobenius norm (i.e. the Euclidean norm of the flattened tensor).
	\item Root-mean-square error (\texttt{-r} flag):
	$$\mathrm{RMSE}(\set{T}, \widetilde{\set{T}}) := \|\set{T} - \widetilde{\set{T}}\| / \sqrt{I_1 \cdots I_N}.$$
	\item Peak signal-to-noise ratio (\texttt{-p} flag): 	$$\mathrm{PSNR}(\set{T}, \widetilde{\set{T}}) := 20 \cdot \log_{10}\left(\frac{\max\{\set{T}\} - \min\{\set{T}\}}{2 \cdot \mathrm{RMSE}(\set{T}, \widetilde{\set{T}})}\right).$$
\end{itemize}

The target specified is then converted to sum of squared errors (SSE) for the algorithm's internal use via the following equivalences:
\begin{equation}
\mathrm{SSE} = \eps^2 \cdot \|\set{T}\|^2 = \mathrm{RMSE}^2 \cdot C =  \left(\frac{\max\{\set{T}\} - \min\{\set{T}\}}{2 \cdot 10^{\mathrm{PSNR}/20}}\right)^2 \cdot C
\end{equation}
where $C$ is the total number of grid points $I_1 \cdots I_N$.

The algorithm consists of three main stages. First, the full non-truncated HOSVD is run on the input data set to yield $N$ orthogonal square factor matrices and an $N$-dimensional core of the same size as the original. The HOSVD core is flattened as a 1D vector of $C$ coefficients, which are then scaled and cast as 64-bit integers. We use C-ordering, i.e. dimensions in the core are traversed from right to left. Conceptually, we handle that sequence of integers as a $C \times 64$ binary matrix $\mat{M}$. Second, a number of that matrix's leftmost columns (the \emph{bit planes}) are compressed without loss, namely the least number such that the overall $l^2$ error falls under a given target. This compression is achieved via run-length encoding (RLE) followed by arithmetic coding (AC). Last, the factor matrices are compressed using a cost-efficient budget criterion. See Algs.~\ref{alg:tthresh} and~\ref{alg:tthresh2} for a pseudocode of our compression pipeline; its individual building blocks are detailed next.

\begin{algorithm}[ht]
\begin{algorithmic}[1]
\State $\set{B} := \set{T}$
\State \emph{// HOSVD transform}
\For{$n = 1, \dots, N$}
	\State $\mat{B}_{(n)} := \mathrm{unfold}(\set{B}, n)$ \emph{// Size $I_n \times (I_1 \cdots \widehat{I_n} \cdots I_N)$}
	\State $\widehat{\mat{B}}_{(n)} := \mat{B}_{(n)} \cdot \mat{B}_{(n)}^T$ \emph{// Symmetric matrix of size $I_n \times I_n$}
	\State $\mat{\Lambda}^{(n)}, \mat{U}^{(n)} = \mathrm{eig}(\widehat{\mat{B}}_{(n)})$ \emph{// Full decomposition; eigenvalues $\mat{\Lambda}^{(n)}$ in non-increasing order}
	\State $\mat{B}_{(n)} := {\mat{U}^{(n)}}^T \cdot \mat{B}_{(n)}$ \emph{// Right part $\mat{\Sigma} \cdot \mat{V}^T$ of the SVD}
	\State $\set{B} := \mathrm{fold}(\mat{B}_{(n)})$ \emph{// Back to original size}
\EndFor
\State \emph{// $\set{B}$ is now the HOSVD core, and $\mat{U}^{(1)}, \dots, \mat{U}^{(N)}$ its factors}

\State $\alpha_b := \mathrm{\textsc{encode}}(\set{B}, s)$ \emph{// See Alg.~\ref{alg:tthresh2}}
\For{$n = 1, \dots, N$}
	\State $\mathrm{\textsc{encode}}(\mat{U}^{(n)}, \alpha_b)$
\EndFor
\end{algorithmic}
\caption{Compress an $N$-dimensional tensor $\set{T}$ of size $I_1 \times \dots \times I_N$ at a prescribed sum of squared errors $s$ using \textsc{tthresh}.}
\label{alg:tthresh}
\end{algorithm}

\begin{algorithm}[!ht]
\begin{algorithmic}[1]
\Function{encode}{$x, \alpha_b, s$}
\State $\set{M} := \emptyset$ \emph{// Mask to record coefficients that have already become significant. It starts out empty}
\State $\mat{M} :=$ binary matrix of size $C \times 64$ containing all elements from $x$, in 64-bit unsigned integer format

\vspace{0.2cm}

\State $\tilde{s} := \|x\|^2$ \emph{// We start with the largest SSE}

\State \emph{// Bit planes from more to less significant}
\For{$p = 63, \dots, 0$}
	\For{$c = 1, \dots, C$}
		\If{$c \in \set{M}$} \emph{// The $c$-th coefficient is already significant}
			\State $\mathrm{encodeBitVerbatim}(\mat{M}[c, p+1])$
		\Else
			\State $\mathrm{encodeBitRLE}(\mat{M}[c, p+1])$
			\If{$\mat{M}[c, p+1] == 1$} \emph{// It becomes significant now}
				\State $\set{M} := \set{M} \cup \{c\}$
			\EndIf
		\EndIf
		\State Update current SSE $\tilde{s}$
		\State Estimate current ratio $\tilde{\alpha}$: the reduction in SSE achieved by the last $c$ bits, divided by the number of bits needed to compress them
		\If{$\mathrm{isCore}(x)$ and $\tilde{s} \le s$}
		\State Exit the two nested loops
		\EndIf
		\If{$\mathrm{isFactor}(x)$ and $\tilde{\alpha} \le \alpha_b$}
		\State Exit the two nested loops
		\EndIf
	\EndFor
\EndFor

\If{$\mathrm{isCore}(x)$}
\State \Return $\tilde{\alpha}$
\EndIf
\EndFunction
\end{algorithmic}
\caption{Encode the decomposition parts obtained in Alg.~\ref{alg:tthresh}. The input $x$ can be either the core $\set{B}$ or a factor $\mat{U}^{(n)}$.}
\label{alg:tthresh2}
\end{algorithm}

\subsection{HOSVD Transform} \label{sec:hosvd_transform}

We use the HOSVD as presented in Sec.~\ref{sec:tucker_compression} to compute orthogonal Tucker factors as the left singular vectors of unfolding matrices. We use 64-bit floating point precision. In general one may directly compute these singular vectors from each unfolding $\mat{B}_{(n)} = \mat{U}^{(n)} \cdot \mat{\Sigma}^{(n)} \cdot {\mat{V}^{(n)}}^T$ in one run of any standard SVD algorithm. In volume compression, however, we usually have $\mat{B}_{(n)} \in \mathbb{R}^{I \times J}$ with $I \ll J$. Since we do not need the $J$ right singular vectors, in such cases it is much more efficient to compute first the matrix $\widehat{\mat{B}}_{(n)} = \mat{B}_{(n)} \cdot \mat{B}_{(n)}^T$ and then obtain all left singular vectors $\mat{U}^{(n)}$ from the full eigenvalue decomposition $\widehat{\mat{B}}_{(n)} = \mat{U}^{(n)} \mat{\Lambda}^{(n)} \mat{U}^{(n)T}$. Since $\widehat{\mat{B}}_{(n)}$ is a real symmetric matrix, its eigenvalue diagonalization always exists and we can use a more efficient specialized solver. The remaining rightmost part of the SVD follows from
\begin{equation}
\mat{\Sigma}^{(n)} {\mat{V}^{(n)}}^T = (\mat{U}^{(n)})^{-1} \mat{B}_{(n)} = {\mat{U}^{(n)}}^T \mat{B}_{(n)}
\end{equation}
as the factor matrix $\mat{U}^{(n)}$ is orthogonal. This process is undertaken $N$ times. In the last iteration we reshape (fold) $\mat{\Sigma}^{(N)} \cdot {\mat{V}^{(N)}}^T$ back into an $N$-dimensional tensor, namely the core $\set{B}$.

\subsection{Bit-plane Coding} \label{sec:bitplane_coding}

Once the Tucker core $\set{B}$ is available we can turn to our coefficient coding scheme. Note that, since no truncation was performed, we have not yet incurred any loss of accuracy other than floating-point round-off errors. Our goal now is to produce an approximate core $\widetilde{\set{B}}$ such that its SSE satisfies

\begin{equation} \label{eq:relative_error}
\mathrm{SSE}(\set{B}, \widetilde{\set{B}}) = \|\set{B} - \widetilde{\set{B}}\|^2 \le s,
\end{equation}
where $s$ is a user-defined bound (recall that, due to factor orthogonality as we saw in Sec.~\ref{sec:tucker_model}, compression error is directly related to the error in the core coefficients). We address this via \emph{bit-plane coding} in the spirit of EZW~\cite{Shapiro:93}, SPIHT~\cite{SP:96}, or EBCOT~\cite{Taubman:00}. We start off by scaling each coefficient's absolute value into a 64-bit unsigned integer

\begin{equation}
\label{eq:casting}
c \mapsto \left \lfloor |c| \cdot 2^{63- \lfloor \log_2(m) \rfloor} \right \rfloor,
\end{equation}
where $m :=  \max_{c \in \set{B}}\{|c|\}$ is the core's largest element (in absolute value). The signs are dealt with separately; see later in this section. Each integer as given by Eq.~\ref{eq:casting} has a decomposition in powers of 2: $2^{63} \cdot c_{63} + ... + 2^0 \cdot c_0$ and defines a row of our binary matrix $\mat{M}$. Since in IEEE 754 every 64-bit floating-point number uses at most 53 significant bits, each row of $\mat{M}$ will have at least 11 zero bits. The basic principle that motivates bit-plane coding is the fact that for any bit plane $p$, all bits in the column $\mat{M}[:, p]$ are equally important. We propose a greedy encoding strategy: we transmit first all bits in the most-significant bit plane $p = 63$. We then move on to the next plane, i.e. $p := p-1$, and repeat. We encode each column from top to bottom and terminate as soon as we fall below the given SSE error tolerance $s$, which usually means that we encode only the top portion of the last column. Note that, instead of error, an alternative stopping criterion based on limiting the compressed file size could be similarly devised and the compression process stopped accordingly.

\begin{figure}[ht]
\centering
\subfigure{\includegraphics[width=0.8\columnwidth]{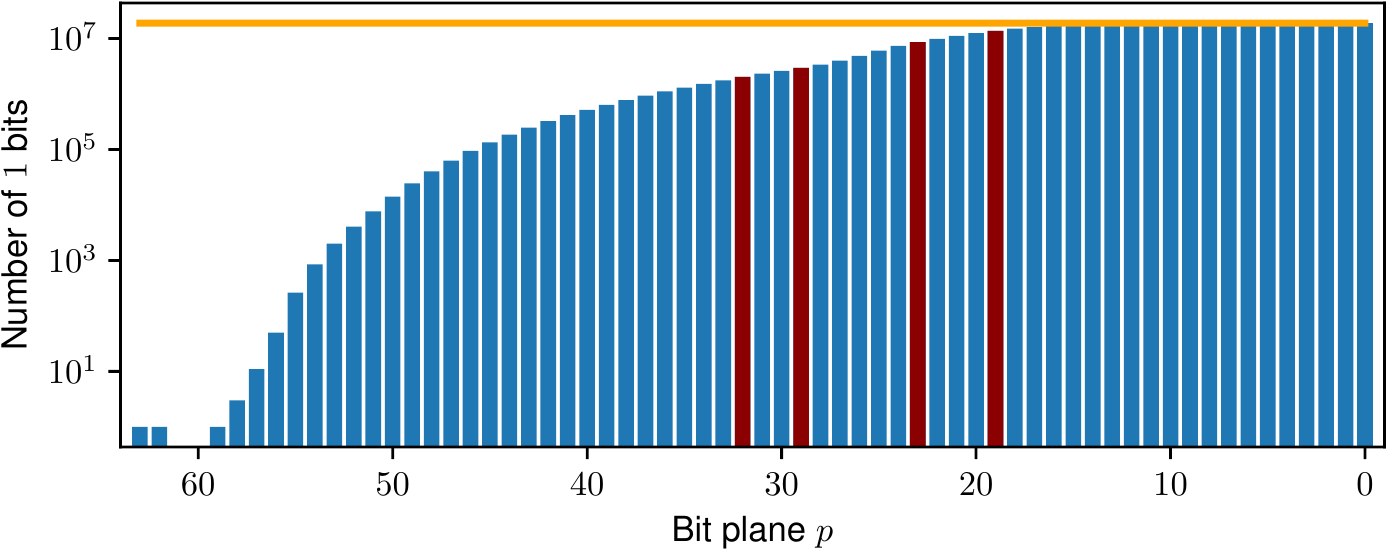}}
\caption{Number of $1$ bits for each bit plane in the HOSVD transform of the Density data set (see Sec.~\ref{sec:results}). The orange line is set at half the total number of core coefficients $C/2$. The threshold plane $P$ needed for 1000:1, 300:1, 100:1, and 50:1 compression is shown for each case from left to right in red.}
\label{fig:one_bars}
\end{figure}

\begin{figure}[ht]
\centering
\subfigure{\includegraphics[width=0.9\columnwidth]{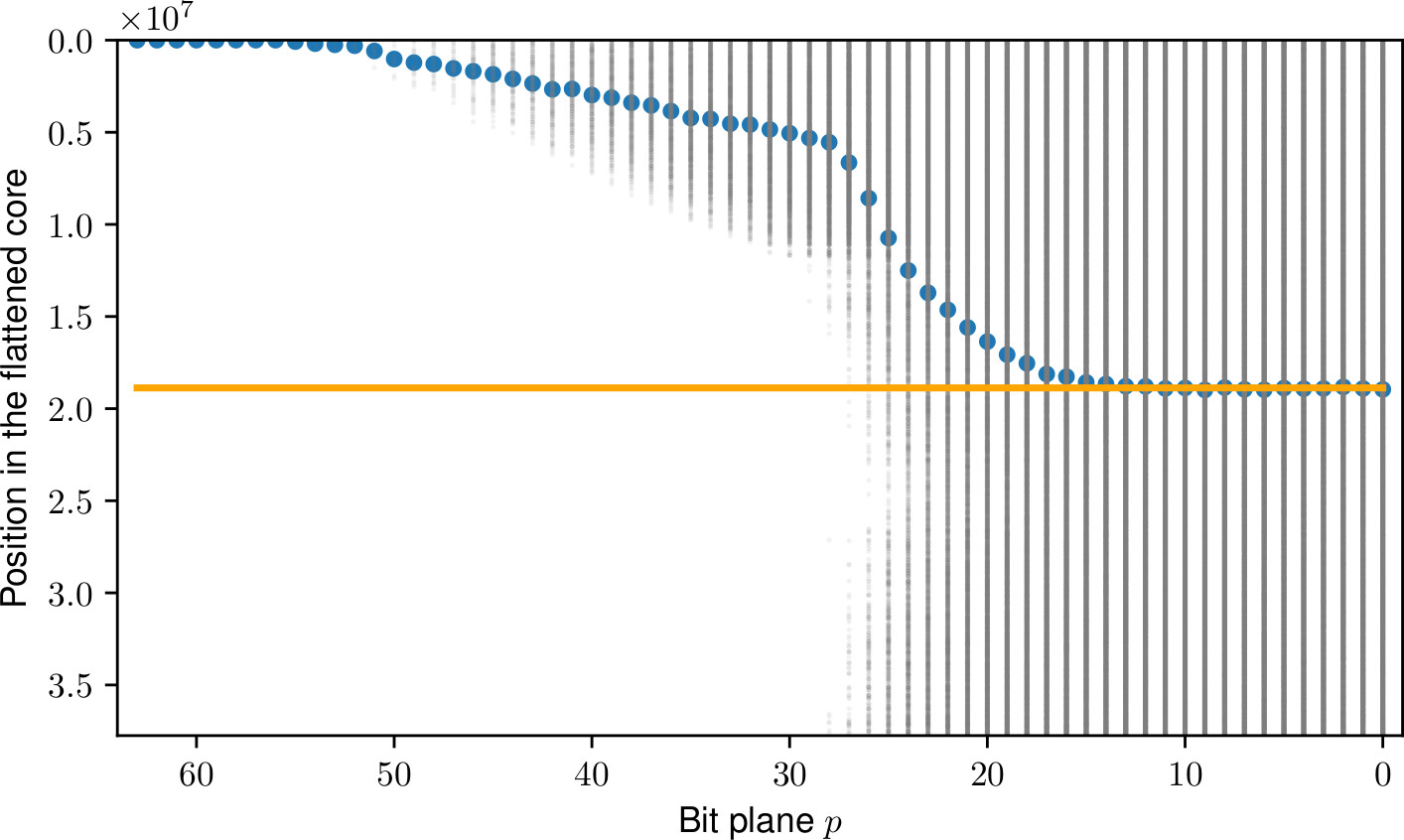}}
\caption{Density plot showing all 1 bits (gray dots) for all bit planes in the flattened HOSVD transform of the Density volume. The center of mass of each set of points is shown as a blue dot. The orange line is set at the center $C/2$.}
\label{fig:core_structure}
\end{figure}

\begin{figure}[ht]
\centering
\subfigure{\includegraphics[width=0.9\columnwidth]{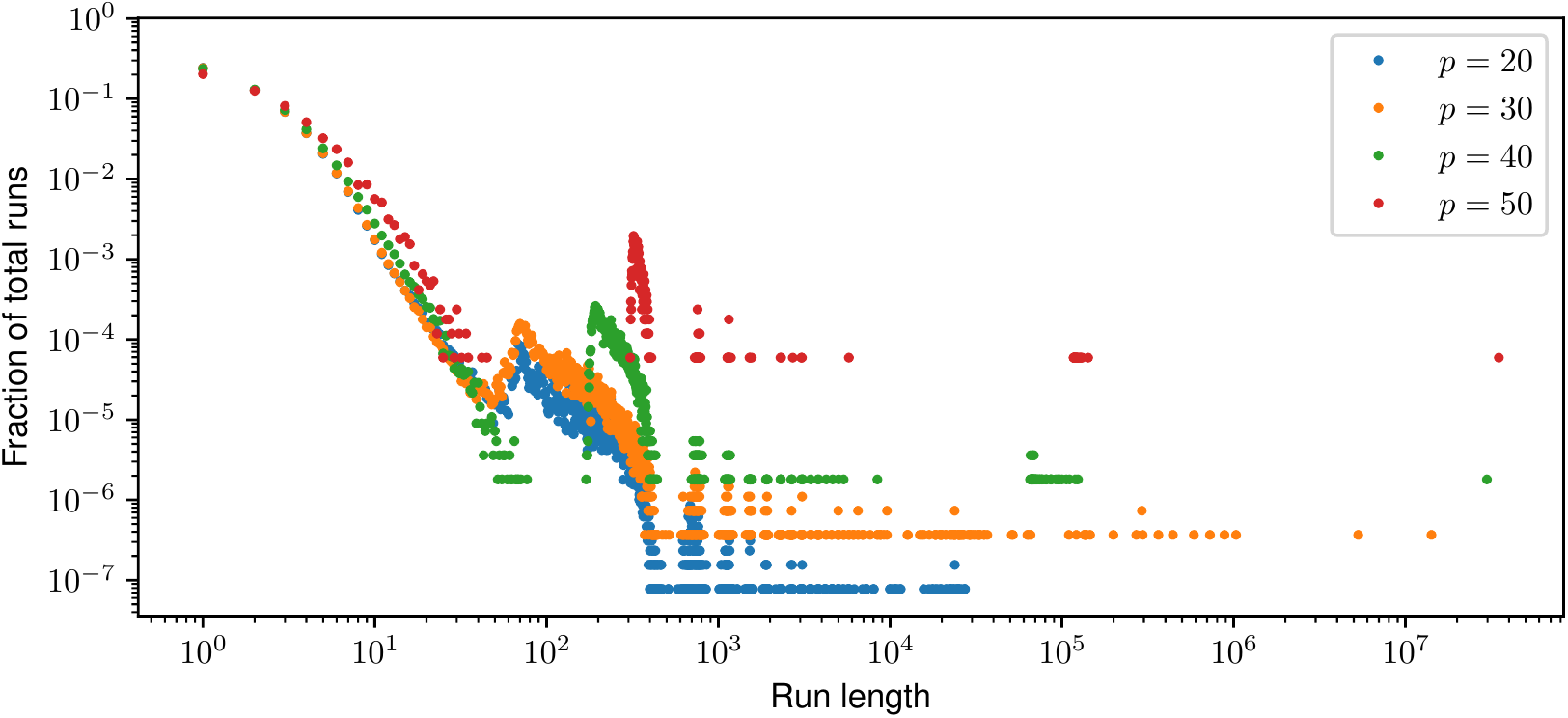}}
\caption{Distribution of run lengths for four different bit planes (Density volume). Note how unbalanced the frequencies are and how they tend to concentrate around a few specific regions along the $x$-axis.}
\label{fig:skipping_stone}
\end{figure}

Since the binary matrix is usually sparse, we initialize it to 0. The error is largest in the beginning and decreases every time a $1$ bit is transmitted. This approach gives the same importance to all bits that lie within the same bit plane, and it ensures that the error it introduces is no larger than the prescribed target SSE. We have chosen a lossless compression strategy to process all selected bits (that is, up to the threshold breakpoint at plane $P$). Statistically, we can expect this to yield high compression ratios thanks to the massive imbalance between the number of 0 and 1 bits in most leading bit planes (see examples in Figs.~\ref{fig:one_bars} and~\ref{fig:core_structure}). Furthermore, long strings of consecutive zero bits (\emph{runs}) happen frequently, and their lengths have a low-entropy distribution. See Fig.~\ref{fig:skipping_stone} for an example; we encode each sequence of $k$ 0-bits that is followed by a 1 (or ends the column) as the integer $k$. For example, the binary string \texttt{01110001} becomes $[$\texttt{1}, \texttt{0}, \texttt{0}, \texttt{3}$]$.

\begin{figure}[ht]
\centering
\subfigure{\includegraphics[width=0.8\columnwidth]{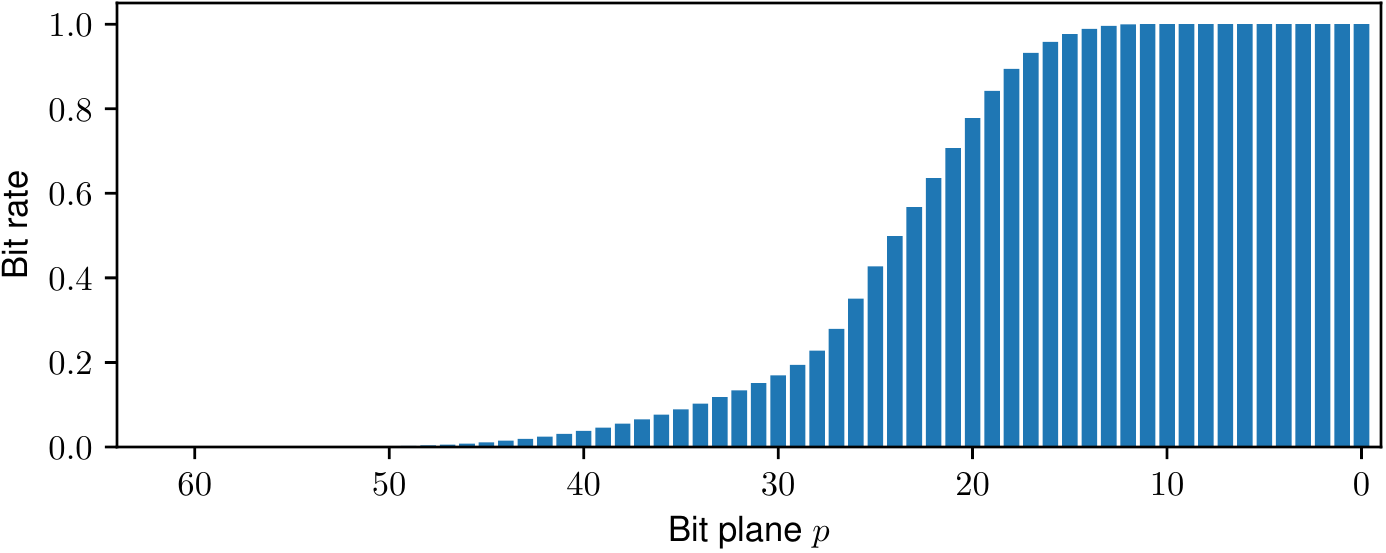}}
\caption{Bit rate that would be achieved by a perfect entropy coder compressing all zero run lengths within each bit plane $p = 63, \dots, 0$ (Density volume). No compression is possible for $p \le 12$.}
\label{fig:optimal_coding}
\end{figure}

Fig.~\ref{fig:optimal_coding} shows theoretical coding performance when storing each bit plane's sequence of zero-run lengths. We plot the bit rate (number of bits after compression, divided by bits before compression) that an ideal entropy coder would need for the given set of integer symbols in the RLE, without accounting for storing a table of frequencies. The bit rate is computed as $\sum_i f_i \cdot \log_2 (n / f_i)$, where $f_i$ counts how many times the $i$-th symbol occurs and $n = \sum_i f_i$ is the total number of symbols to transmit. Modern entropy coders (Huffman and, especially, arithmetic coding) are usually very close to this information-theoretical optimum.

Our statistical analyses on the columns of $\mat{M}$ motivate us to handle each coefficient's \emph{leading bits} (i.e. its leftmost 1 bit and all 0 bits on its left) differently from its \emph{trailing bits} (i.e. 0 or 1 bits that appear to the right of the leftmost 1):

\begin{itemize}
	\item  Leading bits tend to form long runs of zeros with very low entropy along the columns of $\mat{M}$. We compress them without loss via RLE followed by AC.
	\item Trailing bits are close to being uniformly random; RLE+AC cannot compress them well. We thus store them verbatim, which is naturally faster.
\end{itemize}

\begin{figure}[ht]
\centering
\subfigure{\includegraphics[width=0.67\columnwidth]{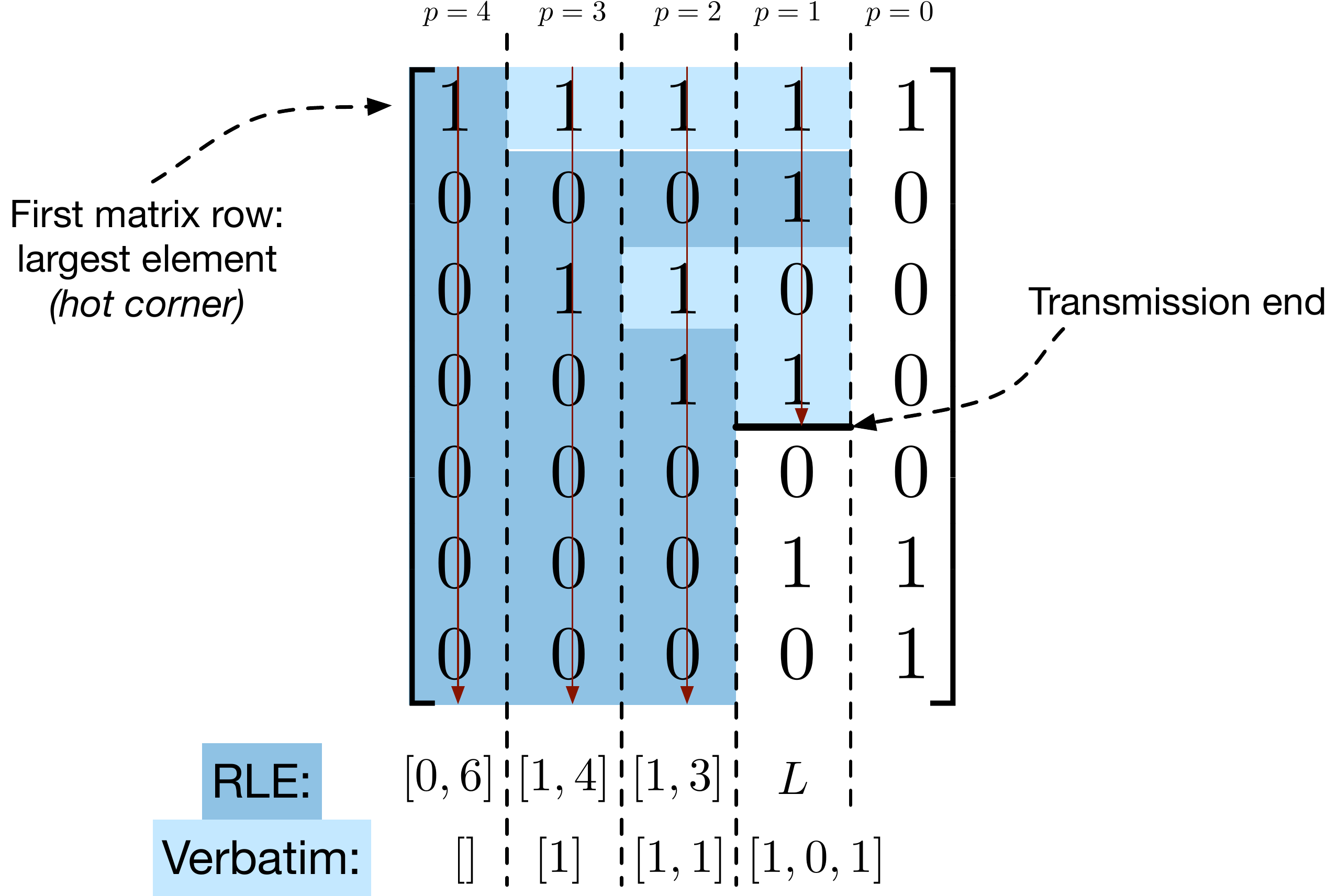}}
\caption{Simplified example coding of 7 coefficients at $P=5$ bits each. Encoded bits are highlighted in dark and light blue; the order is shown by the red arrows, left to right. The transmission was stopped based on a threshold in bit-plane $P=1$. For each coefficient, its leftmost 1 and all leading 0's are compressed using RLE+AC (dark blue), whereas trailing bits are stored verbatim (light blue). The significance mask went from zero members at $p=4$ to three at $p=1$.}
\label{fig:bitplanes}
\end{figure}

Since most planes use a combination of both coding methods, we need an efficient data structure to keep track of leading vs. trailing bits, i.e. a \emph{significance map}. As we work our way from the leftmost $b=63$ towards less significant planes we update a binary mask $\set{M}$ that records all coefficients that have already become significant (their leftmost 1 bit has been encountered). The mask starts empty and gains members progressively. See Fig.~\ref{fig:bitplanes} for a toy illustration.

\paragraph{Remarks}

Recall that this algorithm concerns absolute values only. Like trailing bits, the signs of significant coefficients are close to uniformly random, and we transmit them verbatim as well.

The final compression ratio may vary if one chooses FORTRAN-ordering (left-to-right) instead of C-ordering when flattening the core, since all bit planes will contain different orderings of \texttt{0} and \texttt{1} bits. We found this to influence very little the overall compressed file size in practice.

\subsection{Factor Compression}

The square factors $\{\mat{U}^{(n)}\}_n$ usually account for a small proportion of the overall number of elements in a full HOSVD decomposition, e.g. $(3 \cdot 256^2) / 256^3 \approx 1\%$ for a $256^3$-sized volume. However, this can become a significant overhead if the factors are not compressed as carefully as the core (Sec.~\ref{sec:bitplane_coding}). Although factor matrix compression is an important part of a Tucker-driven compression pipeline, previous related approaches~\cite{SIMAEZGGP:11, SMP:13, BP:15} did not place a particular emphasis on it.

\begin{figure}[ht]
\centering
\subfigure{\includegraphics[width=0.32\columnwidth]{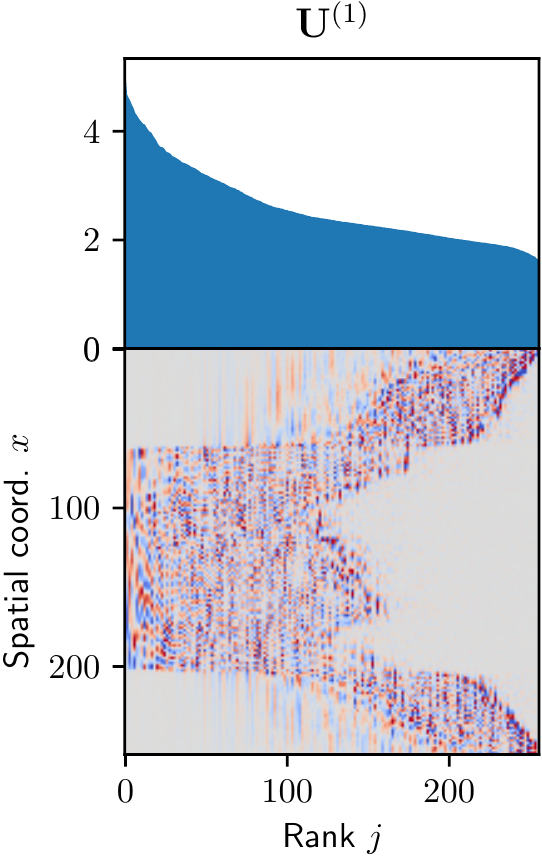}} \hfil
\subfigure{\includegraphics[width=0.32\columnwidth]{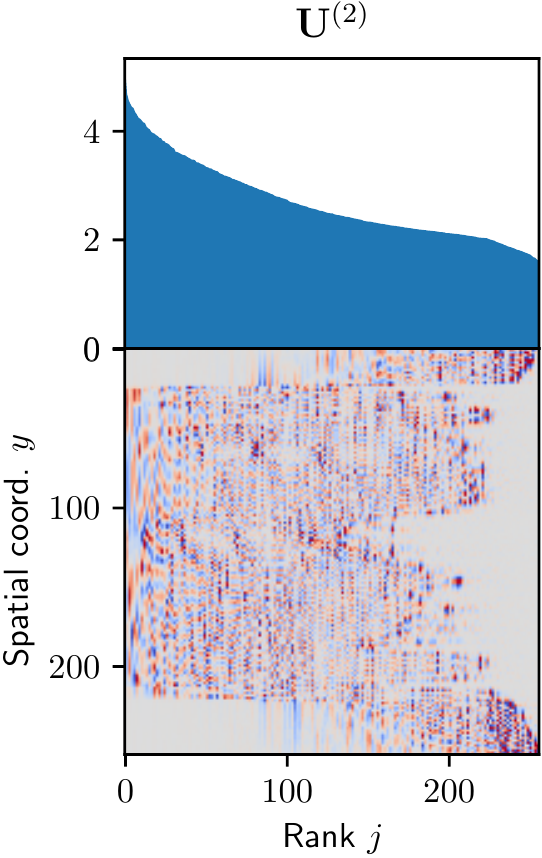}} \hfil
\subfigure{\includegraphics[width=0.32\columnwidth]{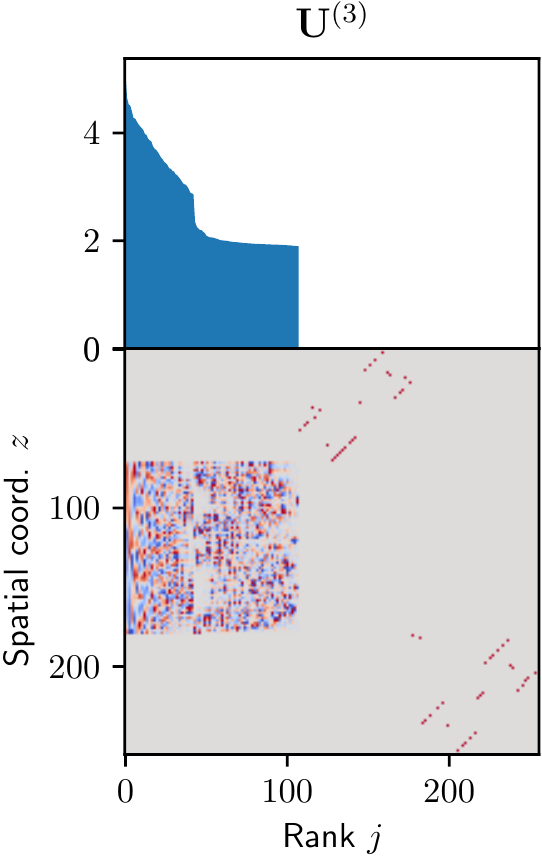}}
\caption{The three factor matrices obtained by decomposing the Engine data set. On top we show, in $\log_{10}$ scale, the corresponding core slice norm $\sigma^{(n)}_j$ of every $j$-th column of each $n$-th factor. Note that, as a consequence of large empty background regions in the volume, many core slices along the third mode are zero.}
\label{fig:slice_norms}
\end{figure}

To encode the factors we essentially reuse the same compression algorithm that we proposed for the core. Nonetheless, two important details deserve special consideration. First, factor matrix columns have vastly different importances: each factor column interacts with one core slice only, and such slices have varying norms (recall Eq.~\ref{eq:non_decreasing}). In practice, those norms are orders of magnitude apart (see Fig.~\ref{fig:slice_norms}), and a proper weighting of our factor columns is in order. Recall (Sec.~\ref{sec:hosvd_transform}) that the factors contain the left singular vectors of an SVD decomposition: $\mat{B}_{(n)} = \mat{U}^{(n)} \cdot \mat{\Sigma}^{(n)} \cdot {\mat{V}^{(n)}}^T$, where $\mat{B}_{(n)}$ is our tensor reconstructed along dimension $n$ only. The matrix $\mat{\Sigma}^{(n)}$ is diagonal and holds the core slice norms $\sigma^{(n)}_1, \dots, \sigma^{(n)}_{I_n}$. Since $\mat{V}^{(n)}$ is orthogonal, any SSE error on $\mat{U}^{(n)} \cdot \mat{\Sigma}^{(n)}$ will produce the same SSE on $\mat{B}_{(n)}$. In other words, in order to control the error that is introduced due to the $n$-th factor we need to compress $\mat{U}^{(n)} \cdot \mat{\Sigma}^{(n)}$. Simply put, we multiply each $j$-th column of $\mat{U}^{(n)}$ by its corresponding core slice norm $\sigma^{(n)}_j$ prior to compression. Since those norms account for a small set of floating point values, we afford to store them explicitly as part as our compression so that the procedure is efficiently reversible for the decompression.

The second question is how many bits we should allocate for each of the factor matrices. Stopping during the same bit plane threshold $P$ that we determined for the core would be a clearly suboptimal choice: even though every factor is just as important as the core for the overall error, they have far fewer elements. Thus, it is reasonable to spend more bits per factor coefficient than we did per core coefficient. We choose a cost-effective criterion that takes into account both compression ratio and quality. Consider the rate-distortion curve obtained by plotting the compression SSE error $s_b$ vs. compressed file size $S_b$ after encoding each core bit $b = 1, \dots, 64 C$. In the beginning we spend zero bits for compression and the error is maximal. The first bits are very \emph{cheap} to encode (they are mostly zeros), yet decrease the error greatly since they belong to the most significant bit planes. In other words, the ratio $\alpha_b := \Delta s_b / \Delta S_b = (s_b-s_{b-1}) / (S_b-S_{b-1})$ is large for $b \ll 64 C$. However, that ratio decreases as more bits are transmitted: average entropy increases, whereas the bit planes lose significance as $p$ decreases. We estimate the ratio $\alpha_b$ achieved at the core transmission breakpoint $b$, and use it as stopping criterion: we halt the encoding of each $n$-th factor at the first bit $b^{(n)}$ such that $\alpha_{b^{(n)}} \le \alpha_b$. By harmonizing all stopping criteria on a single $\alpha_b$, our strategy ensures that a reasonable price is paid in all cases. As we expected, it does result in more bit planes being used than those selected for the core. We also observed that, the smaller the factor matrix, the more bit planes we can generally afford before surpassing $\alpha_b$.

\section{Decompression and Post-processing} \label{sec:compression:decompression}

Decompression follows straightforwardly by inverting the steps described above. HOSVD core bit planes are decompressed in the same order as they were transmitted, the factors are then decompressed, and finally the HOSVD transform is reversed via $N$ TTM products as in Eq.~\ref{eq:tucker}. We speed the TTMs up by detecting and discarding core slices that have become zero during compression. In order to reverse the mixed RLE+AC/verbatim we again use an incrementally-updated mask of significant coefficients similarly to Sec.~\ref{sec:bitplane_coding}. Decompression is significantly faster than compression (see Sec.~\ref{sec:results}) since no covariance and eigenvalue decomposition are needed. After decompression, we apply proper rounding to the core and factors' coefficients. We assume that the residual (i.e. error between the original coefficient and the approximate one) follows an approximately uniform distribution $[0, 2^P-1]$. Instead of simply assuming that the least significant bits $p < P$ of a coefficient are zero, we take the expected value $2^{P-1}$.

\subsection*{Compression-domain Resampling}

Thanks to multilinearity, filtering operations on compressed tensors can be efficiently performed via convolution on their factor matrices; see e.g.~\cite{KK:07, BSP:18}. Separable filters $\set{F} = \vec{u}^{(1)} \otimes \dots \otimes \vec{u}^{(N)}$ are particularly straightforward to apply:
\begin{equation}
\set{T}*\set{F} = \set{B} \times_1 (\mat{U}^{(1)}*\vec{u}^{(1)}) \times_2 \dots \times_N (\mat{U}^{(N)}*\vec{u}^{(N)})
\end{equation}
where $\mat{U}^{(n)}*\vec{u}^{(n)}$ denotes column-wise convolution between a matrix and a column vector. In other words, each row of the $n$-th factor becomes a linear combination of its neighboring rows, weighted by the vector $\vec{u}^{(n)}$. This operation has a negligible cost compared to the decompression, which has to be performed anyway for visualization. Following this principle, we have implemented three options for compressed-domain decimation:

\begin{itemize}
	\item Downsampling: we simply select an evenly spaced subset of the factor rows and discard the rest.
	\item Box filtering: we average consecutive rows together.
	\item Separable Lanczos-2: we convolve column-wise the factors with a 1D Lanczos kernel prior to subselection. We use the 3-lobed kernel, i.e. 5 samples with window parameter equal to 2:
	$$u(x) = \begin{cases}
	\mathrm{sinc}(x) \cdot \mathrm{sinc}(x/2) & \mbox{if } -2 < x < 2 \\ 
	0 & \mbox{otherwise}
	\end{cases}$$
	with $x = \{-2, 1, 0, 1, 2\}$, where $\mathrm{sinc}(x) := \frac{\sin(\pi x)}{\pi x}$.
\end{itemize}

In our implementation the user can specify index ranges and strides via NumPy-style notation. Immediate applications include previewing, subvolume selection and slicing, reversing dimensions, frame-by-frame visualization in time-dependent data, etc. In all these cases the Tucker core remains unchanged, so the filtering and downsampling asymptotic costs amount to only $O(N I^2 \log I)$ operations for the column-wise factor convolution where $I := \max\{I_1, \dots, I_N\}$.

\section{Results} \label{sec:results}

We tested the proposed method with 12 integer and floating-point volume data sets, along with two time-varying volumes (all details and sources are shown in Tab.~\ref{tab:datasets}). We use Eigen 3.2.9 for matrix manipulations, products and eigenvalue decomposition, more specifically its \texttt{SelfAdjointEigenSolver} class for symmetric real matrices. We used a 4-core Intel i7-4810MQ CPU with 2.80GHz and 4GB RAM. All renderings were generated via volume ray casting in ParaView~\cite{paraview}.

\begin{table*}[ht]
\centering
\caption[Data set description]{The 14 data sets tested in this paper.}
\resizebox{2\columnwidth}{!}{
\begin{tabular}{ccccc}
\textbf{Name} & \textbf{Dimensions} & \textbf{Type} & \textbf{Size} & \textbf{Source} \\
\hhline{=====}
``Foot'', ``Engine'' & $256 \times 256 \times 256$ & 8-bit unsigned int & 16 MB & \specialcell{The Volume Library \cite{iapr}} \\
\hline
\specialcell{``Teapot''} & $256 \times 256 \times 178$ & 8-bit unsigned int & 11.1 MB & \specialcell{The Volume Library \cite{iapr}} \\
\hline
\specialcell{``Isotropic-coarse'', \\ ``Isotropic-fine'', ``Channel'', \\ ``MHD'', ``Mixing''} & $512 \times 512 \times 512$ & 32-bit float & 512 MB & \specialcell{All available pressure fields from the \\ Johns Hopkins Turbulence Database \cite{jhtd}} \\
\hline
\specialcell{``Viscosity'', ``Density''} & $384 \times 384 \times 256$ & 64-bit float & 288 MB & \specialcell{Lawrence Livermore National Laboratory \\ (Miranda simulation \cite{CC:06})} \\
\hline
``U'' & $288 \times 192 \times 28$ & 64-bit float & 11.8 MB & \specialcell{National Center for Atmospheric Research \\ (Community Earth System Model \cite{cesm})} \\
\hline
``Jet-u'' & $400 \times 250 \times 200$ & 64-bit float & 152.6 MB & \specialcell{Sandia National Laboratories \\ (S3D simulation \cite{GGYC:11})} \\
\hline
\specialcell{``Isotropic-fine-time''} & $64 \times 64 \times 64 \times 64$ & 32-bit float & 64 MB & \specialcell{Time-varying version of \\ the Isotropic-fine (third row)} \\
\hline
\specialcell{``Hurricane''} & $50 \times 50 \times 91 \times 48$ & 32-bit float & 42 MB & SciVis 2004 Contest Data Set, QVAPOR field \\
\hline
\end{tabular}
}
\label{tab:datasets}
\end{table*}

We have measured the compression performance of \textsc{tthresh} against four state-of-the-art algorithms:

\begin{itemize}
	\item Tucker rank truncation and fixed core quantization~\cite{SMP:13} (our own implementation). We use 8 and 32 bits for core and factor coefficients, respectively, and label this algorithm as \textsc{trunc}.
	\item \textsc{zfp}~\cite{Lindstrom:14} (version 0.5.4 as implemented in~\cite{zfp}). We use its fixed accuracy mode (which usually yields the best compression rates), serial execution mode, and vary its absolute error tolerance (\texttt{-a}).
	\item \textsc{sz}~\cite{DC:16} (version 2.0.1.0 as implemented in~\cite{sz}). We use the relative error bound mode and vary accordingly the relative bound ratio parameter (\texttt{relBoundRatio}).
	\item \textsc{sq}~\cite{IKK:12} (our own implementation). We vary the absolute error tolerance and stream the output through the \textsc{lzma} lossless compressor as advised in the original paper.
\end{itemize}

\begin{figure*}[ht]
\centering
\subfigure{\includegraphics[width=0.5\columnwidth]{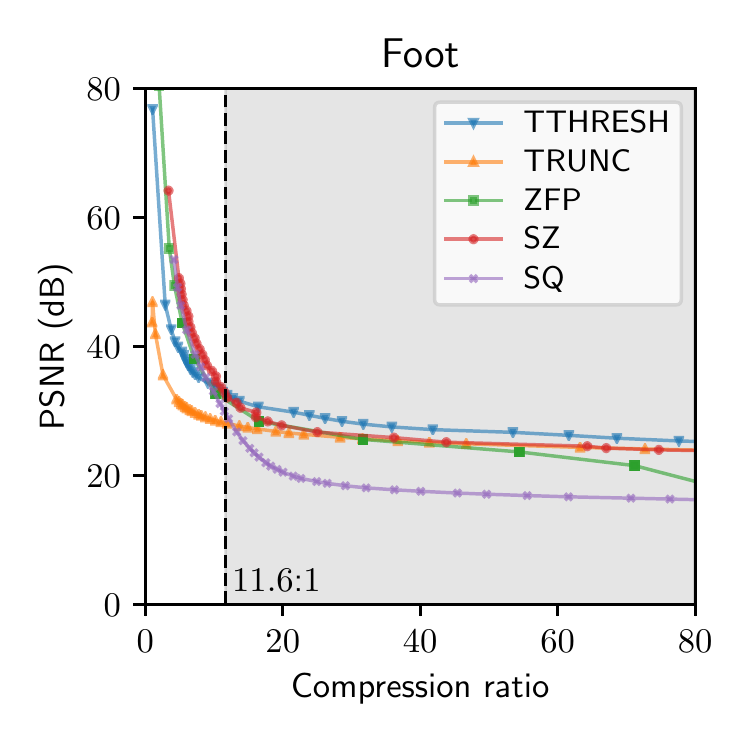}} \hfil
\subfigure{\includegraphics[width=0.5\columnwidth]{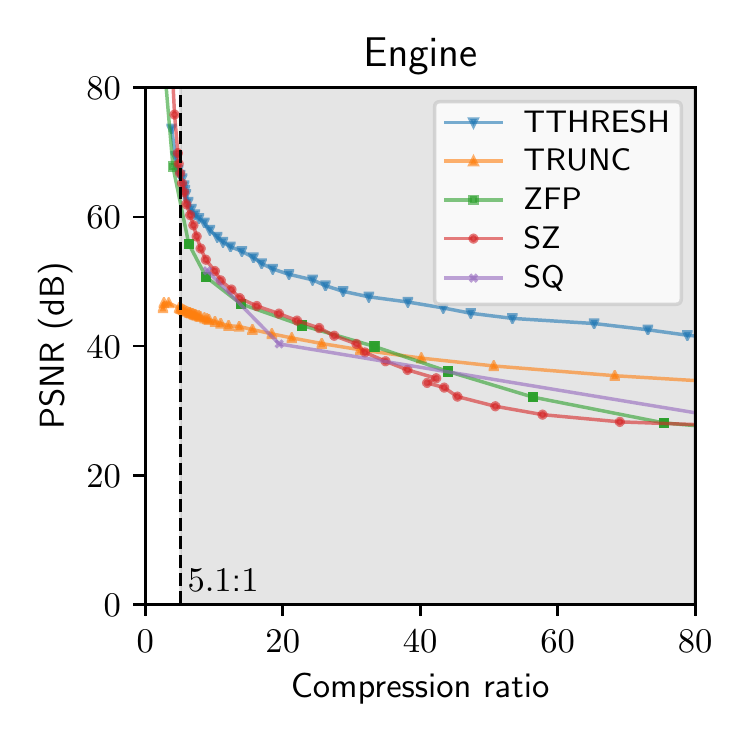}} \hfil
\subfigure{\includegraphics[width=0.5\columnwidth]{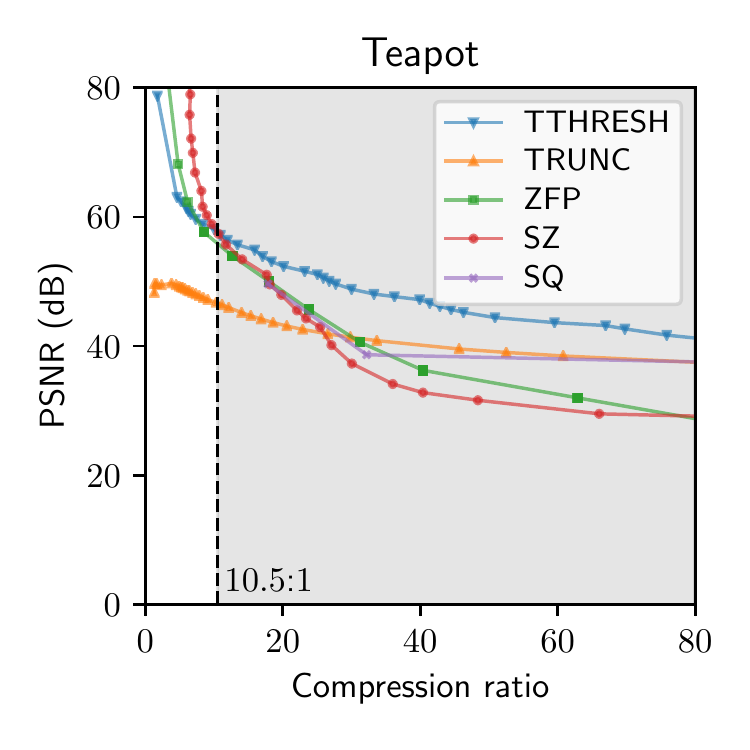}} \hfil
\subfigure{\includegraphics[width=0.5\columnwidth]{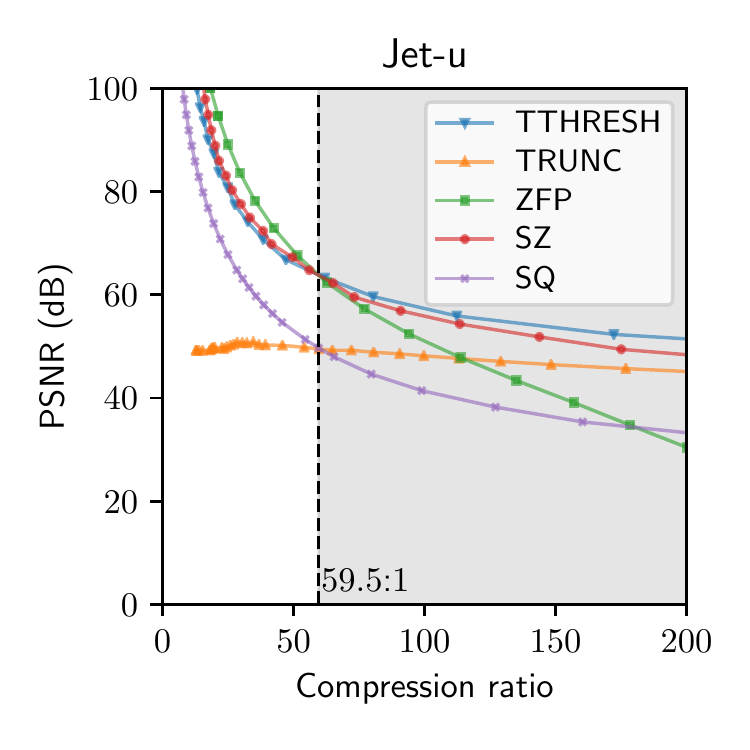}} \vspace{-0.6cm} \\
\subfigure{\includegraphics[width=0.5\columnwidth]{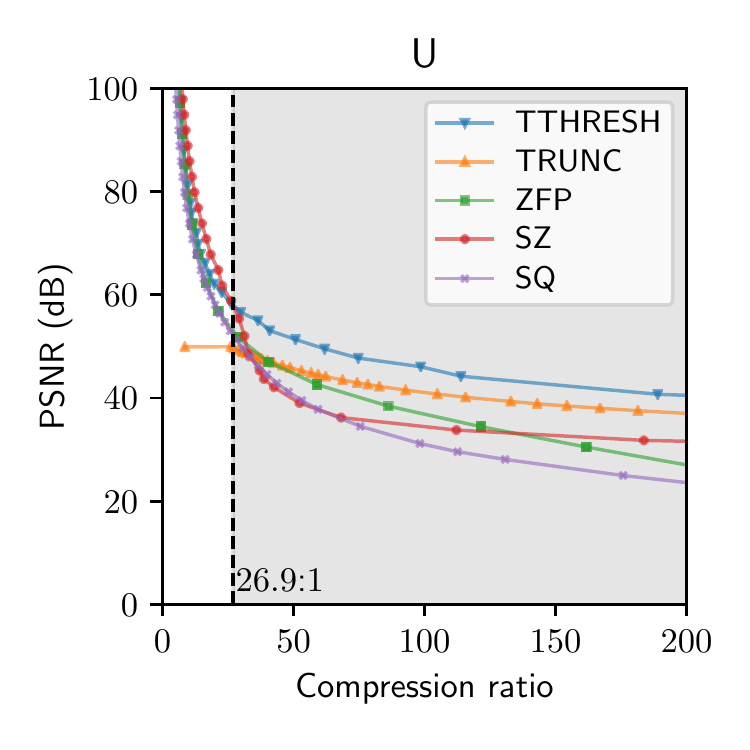}} \hfil
\subfigure{\includegraphics[width=0.5\columnwidth]{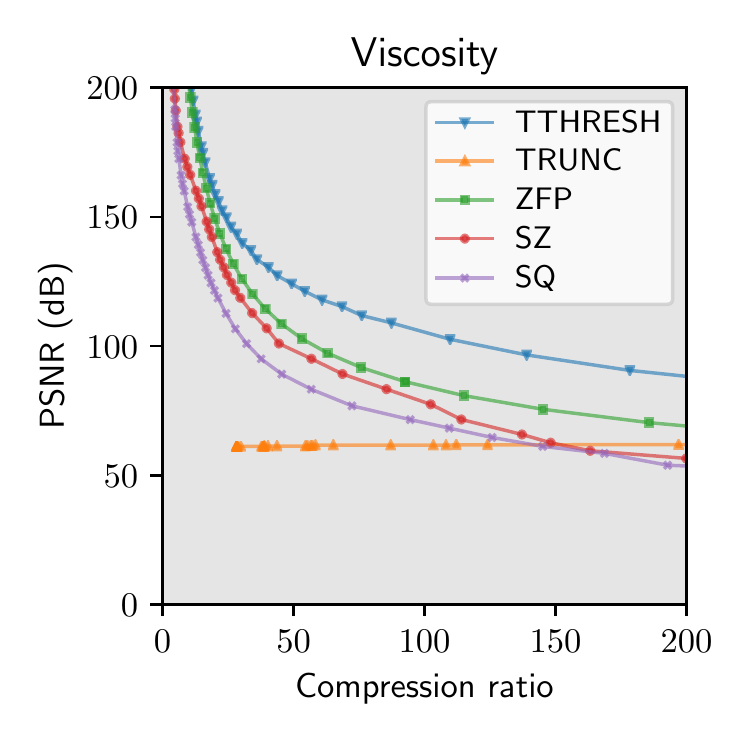}} \hfil
\subfigure{\includegraphics[width=0.5\columnwidth]{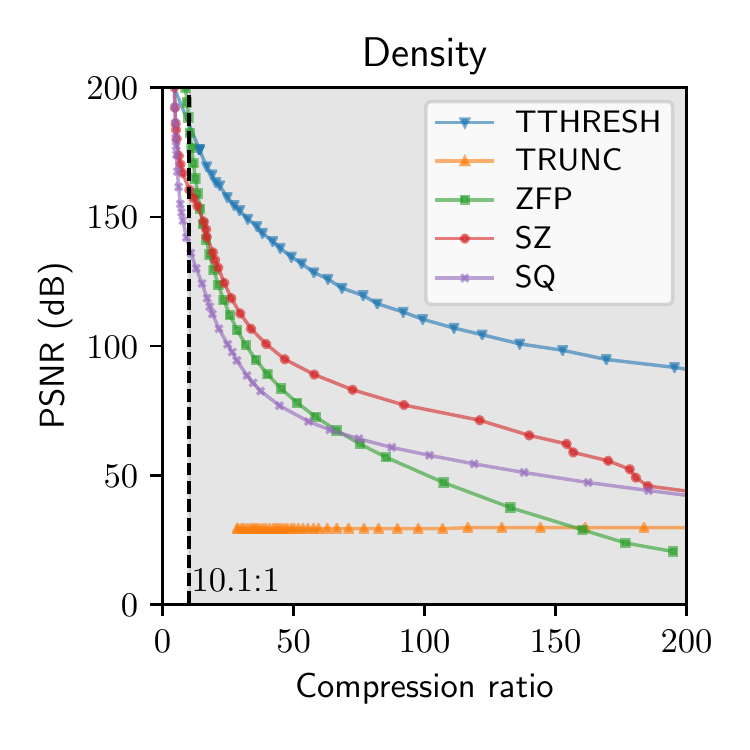}} \hfil
\subfigure{\includegraphics[width=0.5\columnwidth]{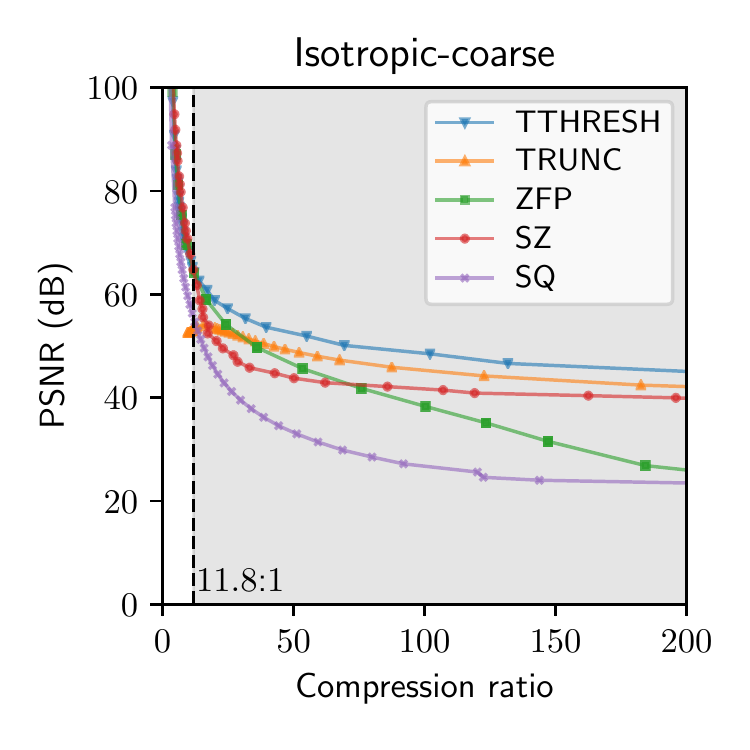}} \vspace{-0.6cm} \\
\subfigure{\includegraphics[width=0.5\columnwidth]{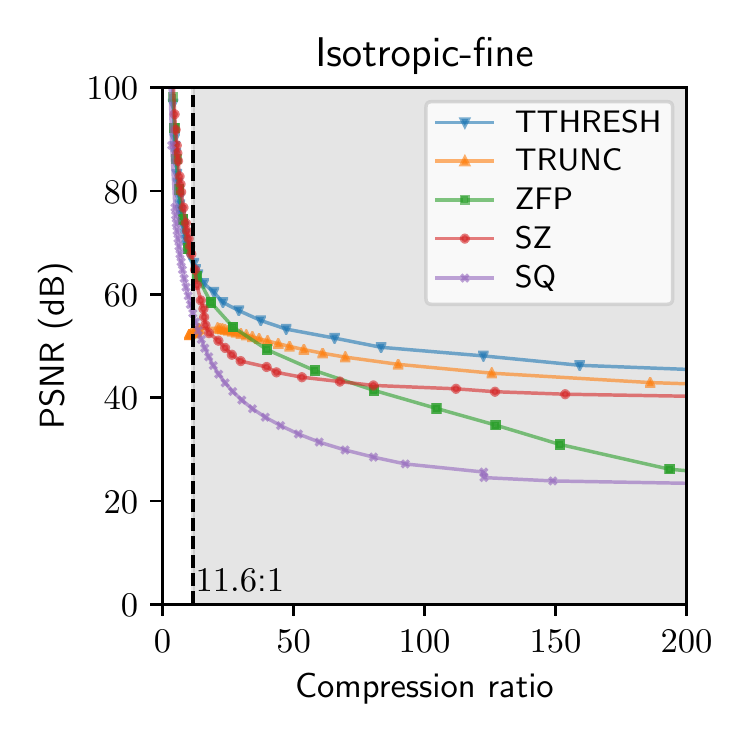}} \hfil
\subfigure{\includegraphics[width=0.5\columnwidth]{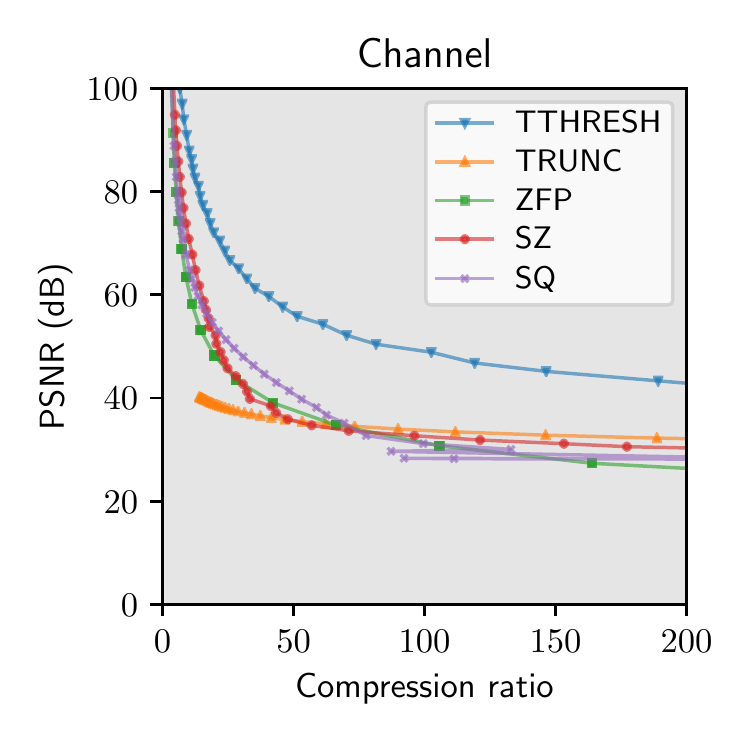}} \hfil
\subfigure{\includegraphics[width=0.5\columnwidth]{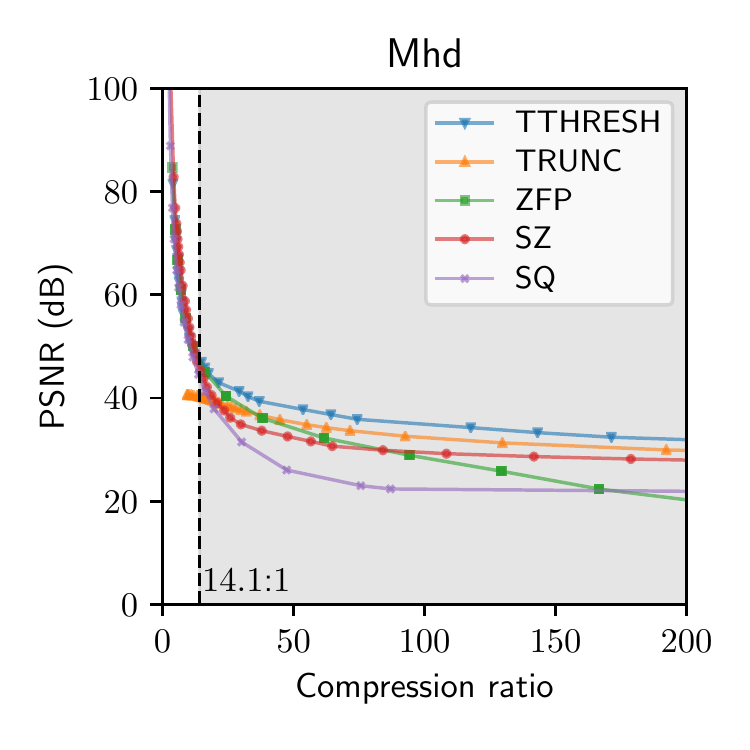}} \hfil
\subfigure{\includegraphics[width=0.5\columnwidth]{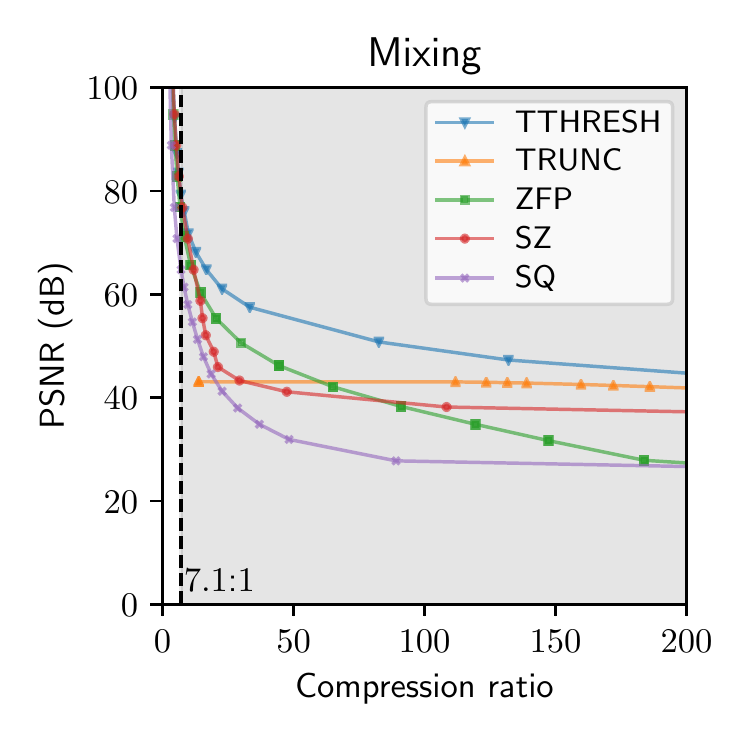}} \\
\caption[PSNR quality curves for several compressors]{Compression quality curves (higher is better) for our method compared to \textsc{trunc}, \textsc{zfp}, \textsc{sz}, and \textsc{sq} over 12 example volumes and varying compression ratios (up to 80:1 for integer data, and 200:1 for floating-point data). We show in gray all ratios where \textsc{tthresh} offers the highest PSNR among all compressors.}
\label{fig:tthresh_plots}
\end{figure*}

\begin{figure}[ht]
\centering
\subfigure{\includegraphics[width=0.49\columnwidth]{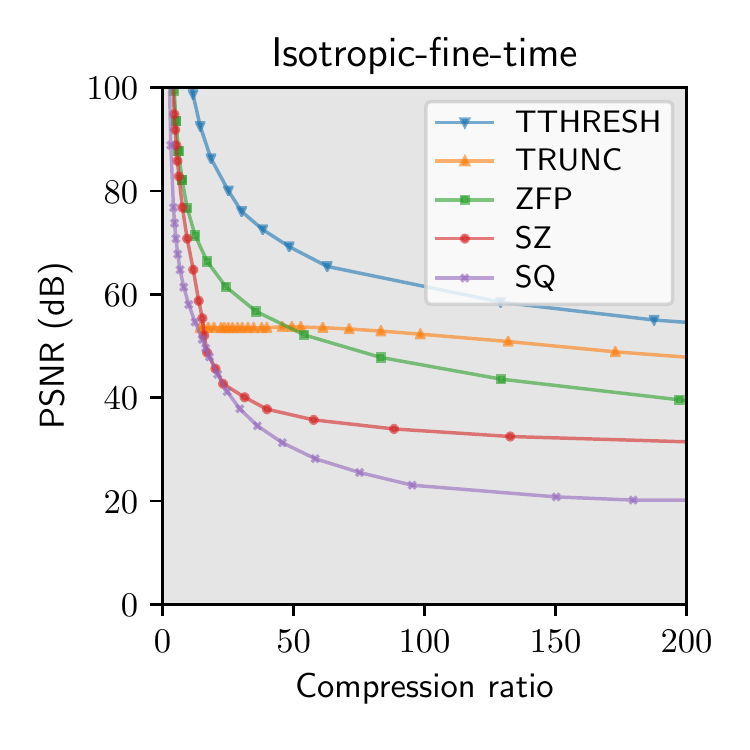}} \hfil
\subfigure{\includegraphics[width=0.49\columnwidth]{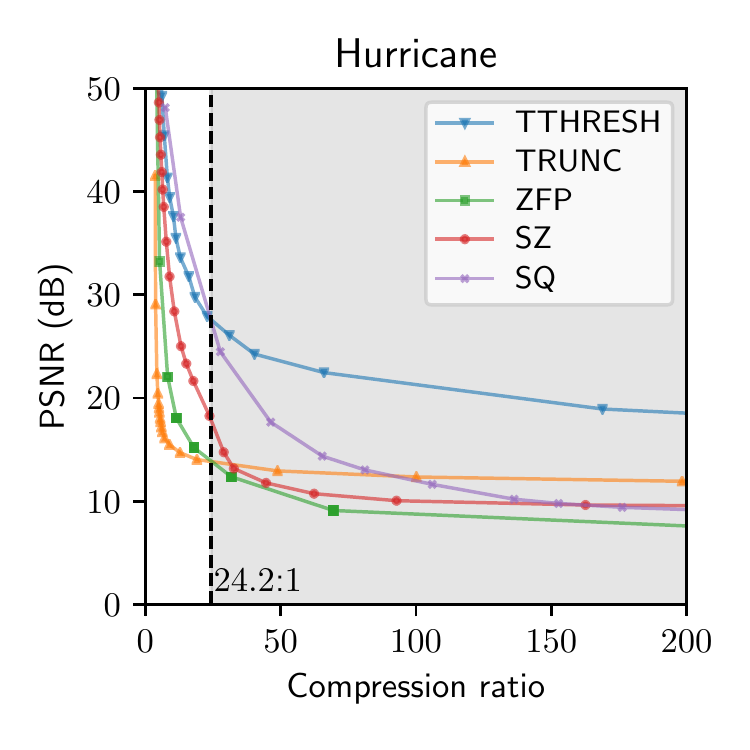}}
\caption{Compression quality curves for two time-varying volumes; see also Fig.~\ref{fig:tthresh_plots}.}
\label{fig:tthresh_plots2}
\end{figure}

All codes were compiled with \texttt{g++} at maximum optimization (\texttt{-O3} flag). Since \textsc{sz} does not readily support integer data types, we first cast all 8-bit volumes to 64-bit floats; we measure compression ratios w.r.t. the original data for all five compressors. Figs.~\ref{fig:tthresh_plots} and~\ref{fig:tthresh_plots2} show the resulting error curves in terms of PSNR vs. compression ratio over all sample data sets. We observe a recurring pattern from lower to higher compression ratio: our proposed algorithm performs similarly (sometimes worse) than other methods for lower ratios, up to a tipping point after which it is better by a widening margin. Although this point can vary significantly, the general behavior is consistent across all data sets we tested. We argue that the usual rates at which \textsc{tthresh} performs best are the most adequate for visualization purposes. 

\begin{figure*}[ht]
\centering
\subfigure{\includegraphics[width=0.47\columnwidth]{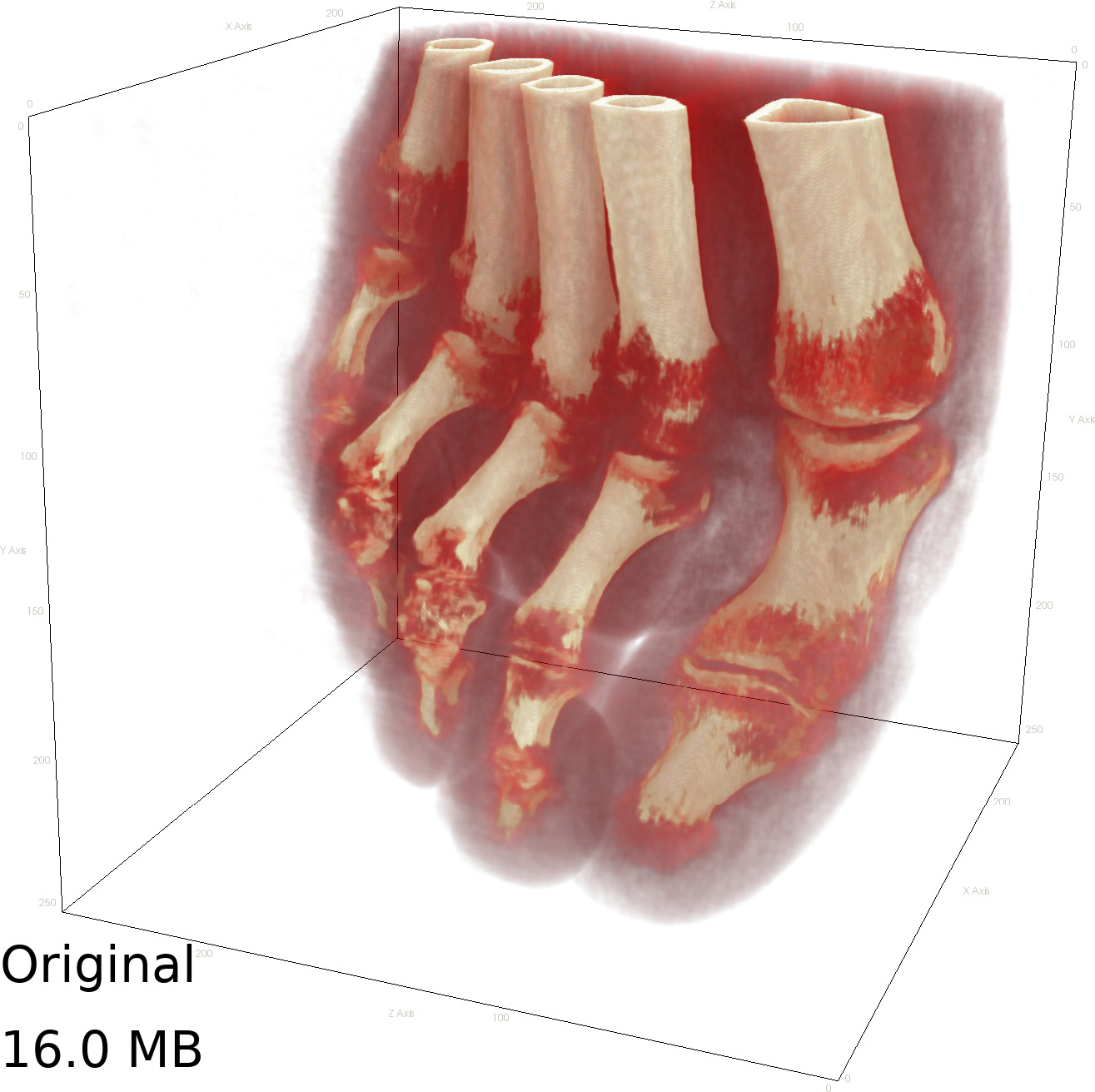}} \hfil
\subfigure{\includegraphics[width=0.47\columnwidth]{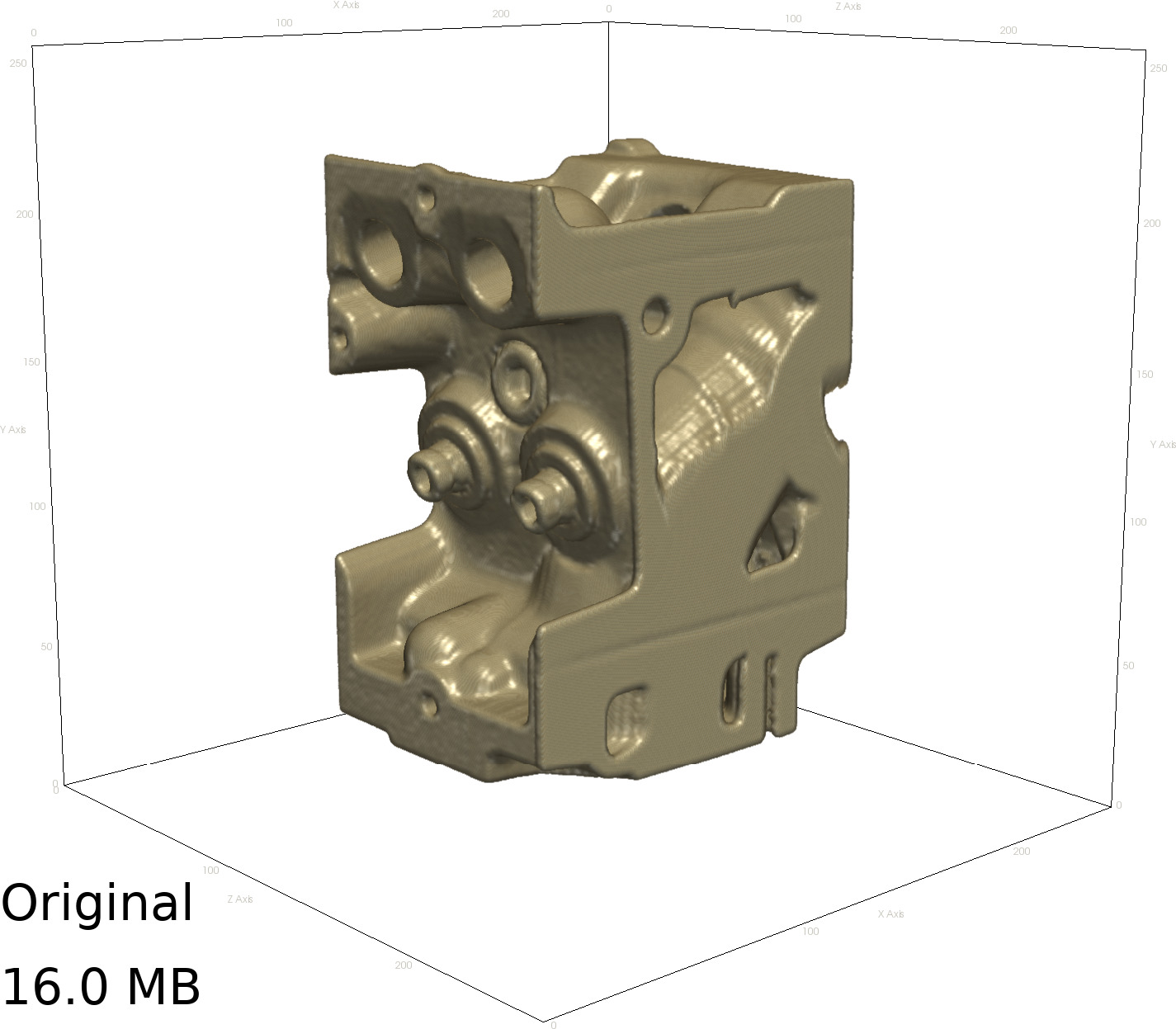}} \hfil
\subfigure{\includegraphics[width=0.47\columnwidth]{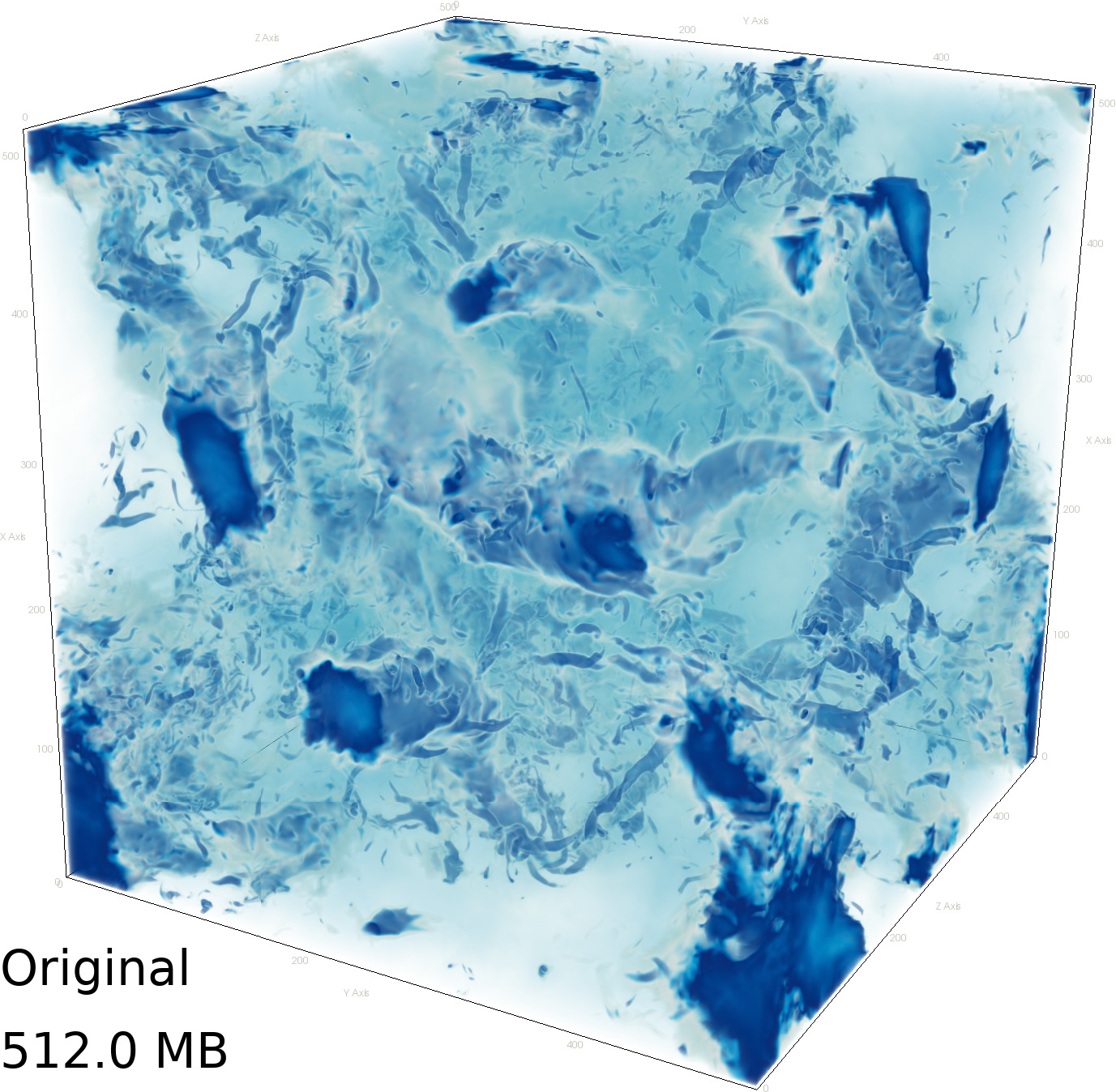}} \hfil
\subfigure{\includegraphics[width=0.47\columnwidth]{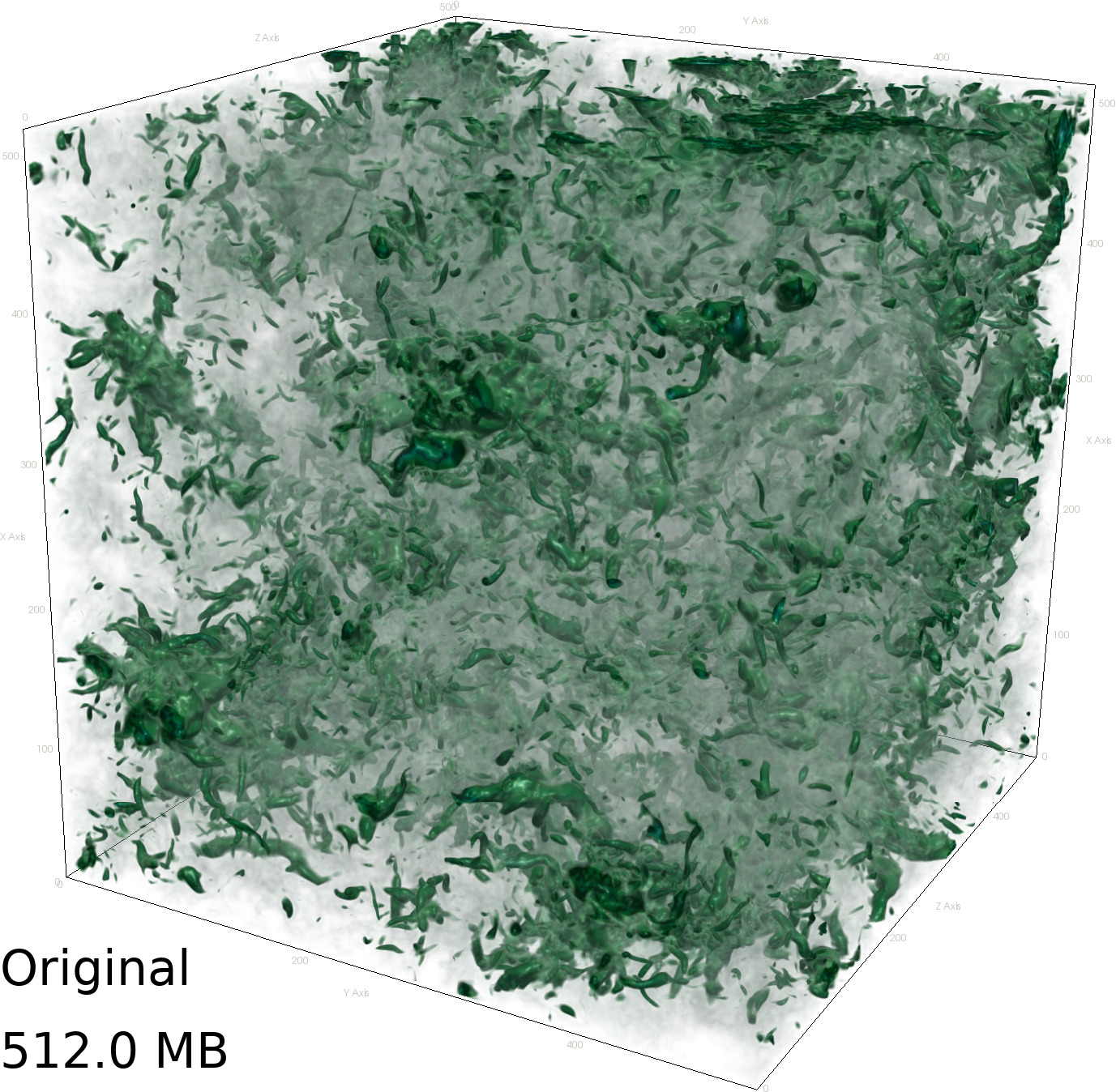}} \\
\subfigure{\includegraphics[width=0.47\columnwidth]{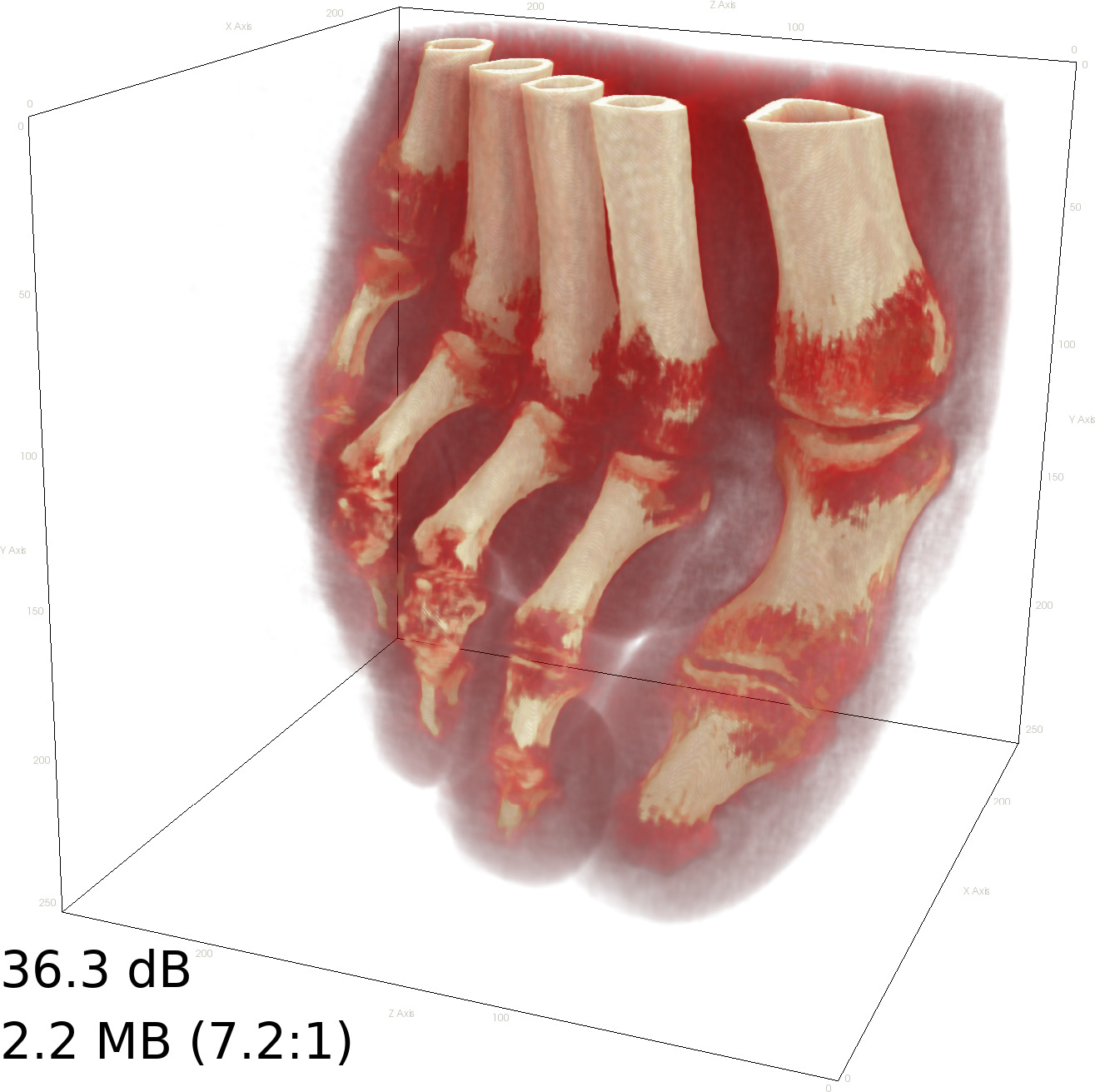}} \hfil
\subfigure{\includegraphics[width=0.47\columnwidth]{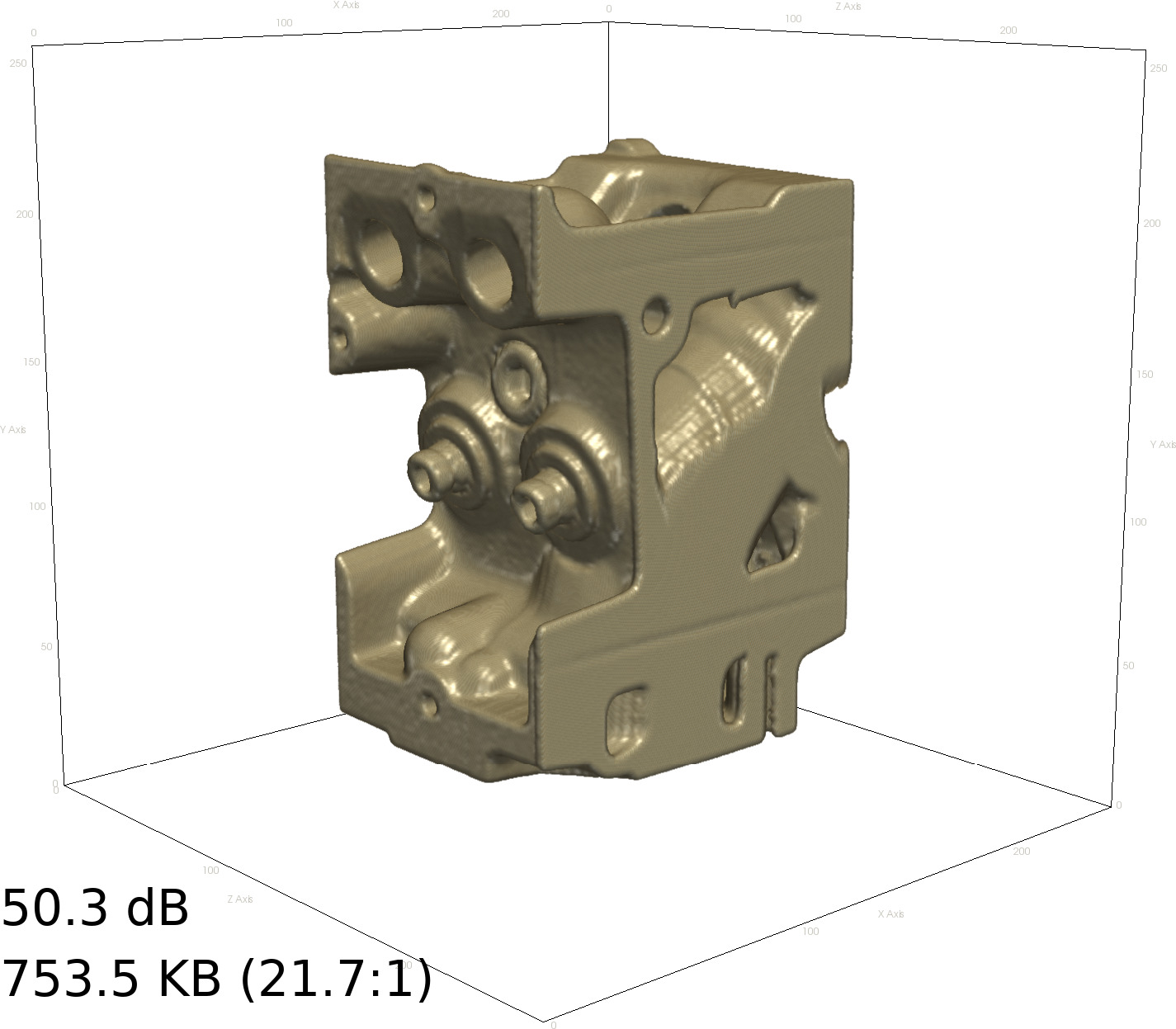}} \hfil
\subfigure{\includegraphics[width=0.47\columnwidth]{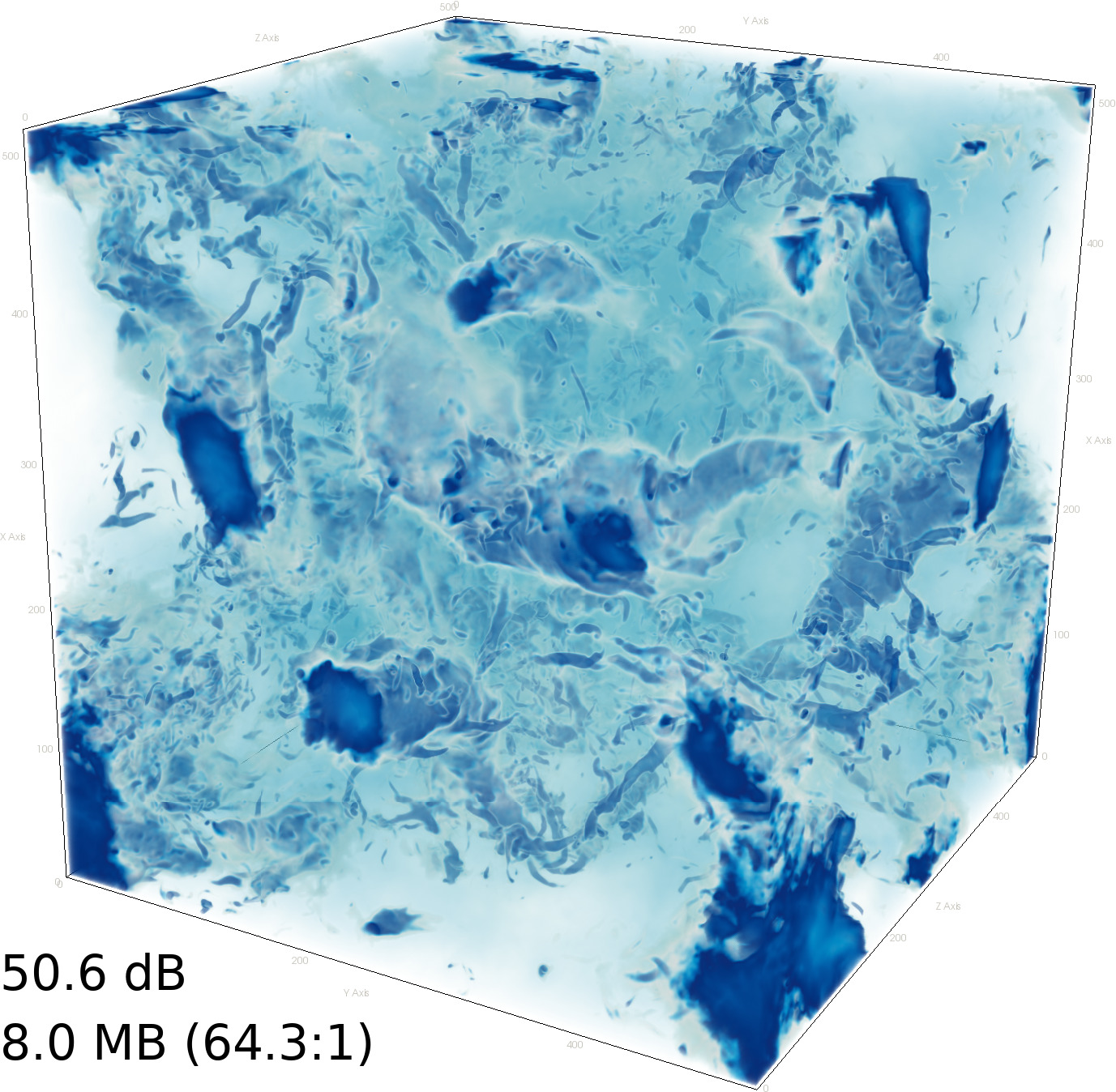}} \hfil
\subfigure{\includegraphics[width=0.47\columnwidth]{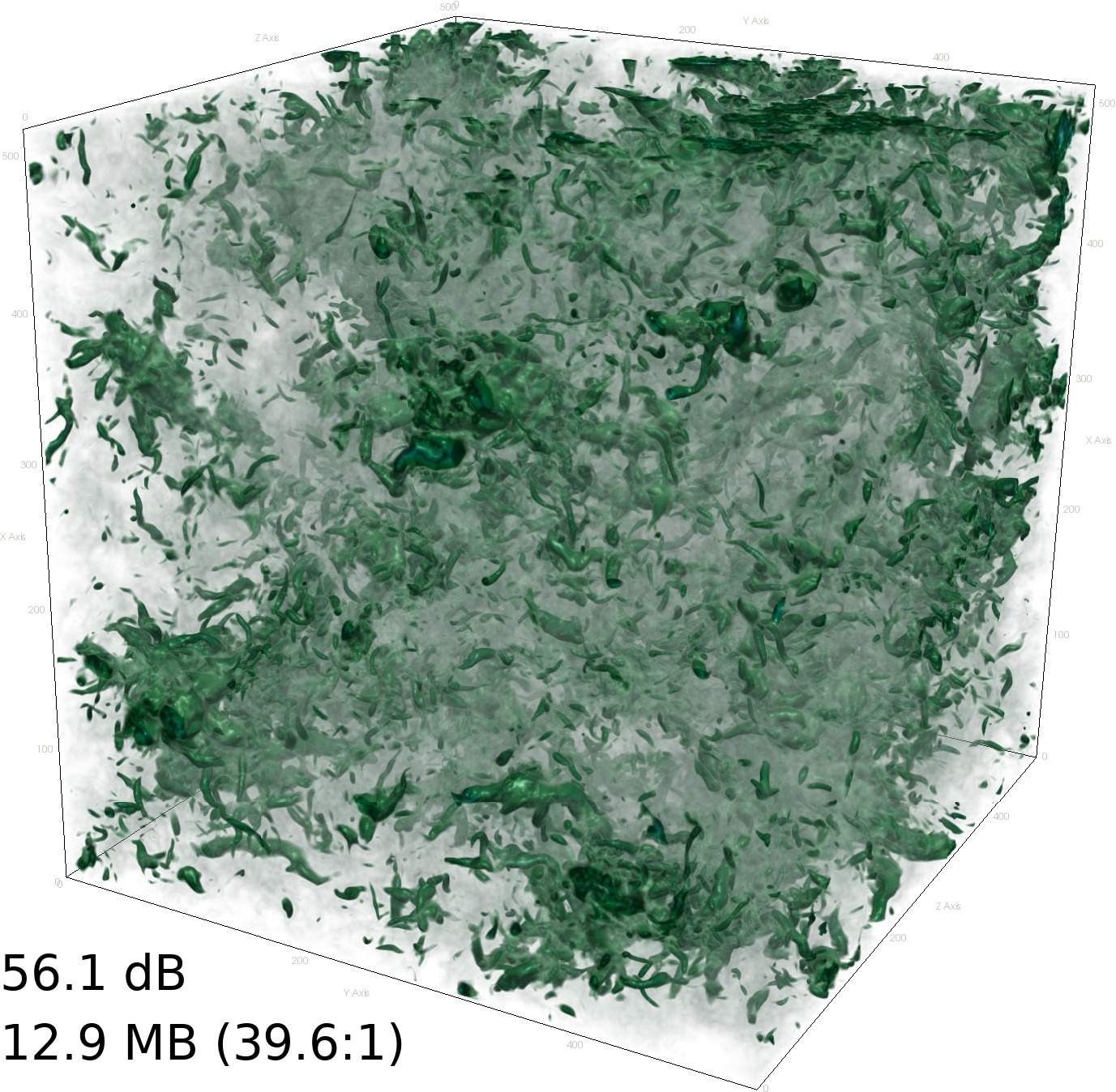}} \\
\subfigure{\includegraphics[width=0.47\columnwidth]{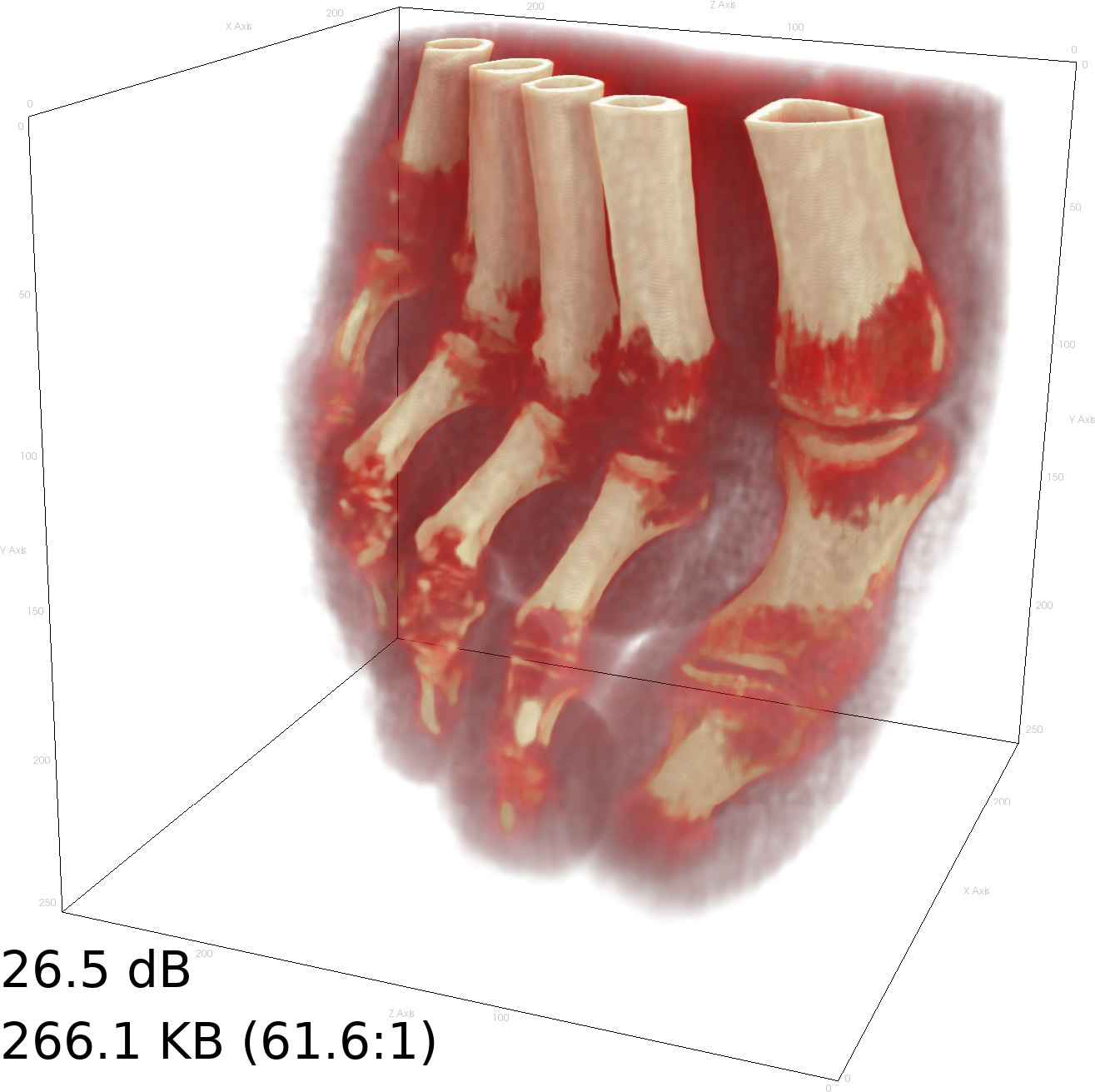}} \hfil
\subfigure{\includegraphics[width=0.47\columnwidth]{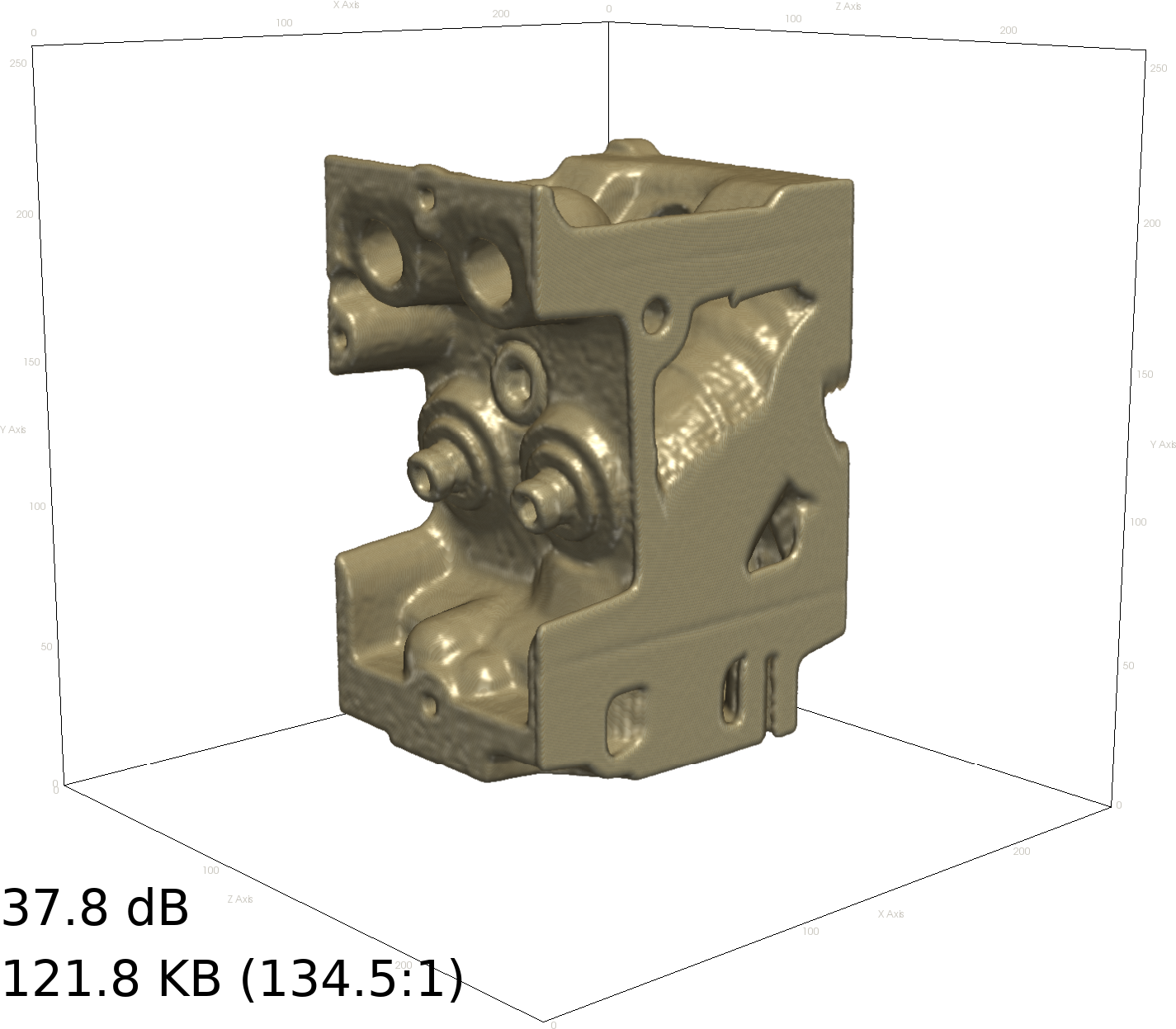}} \hfil
\subfigure{\includegraphics[width=0.47\columnwidth]{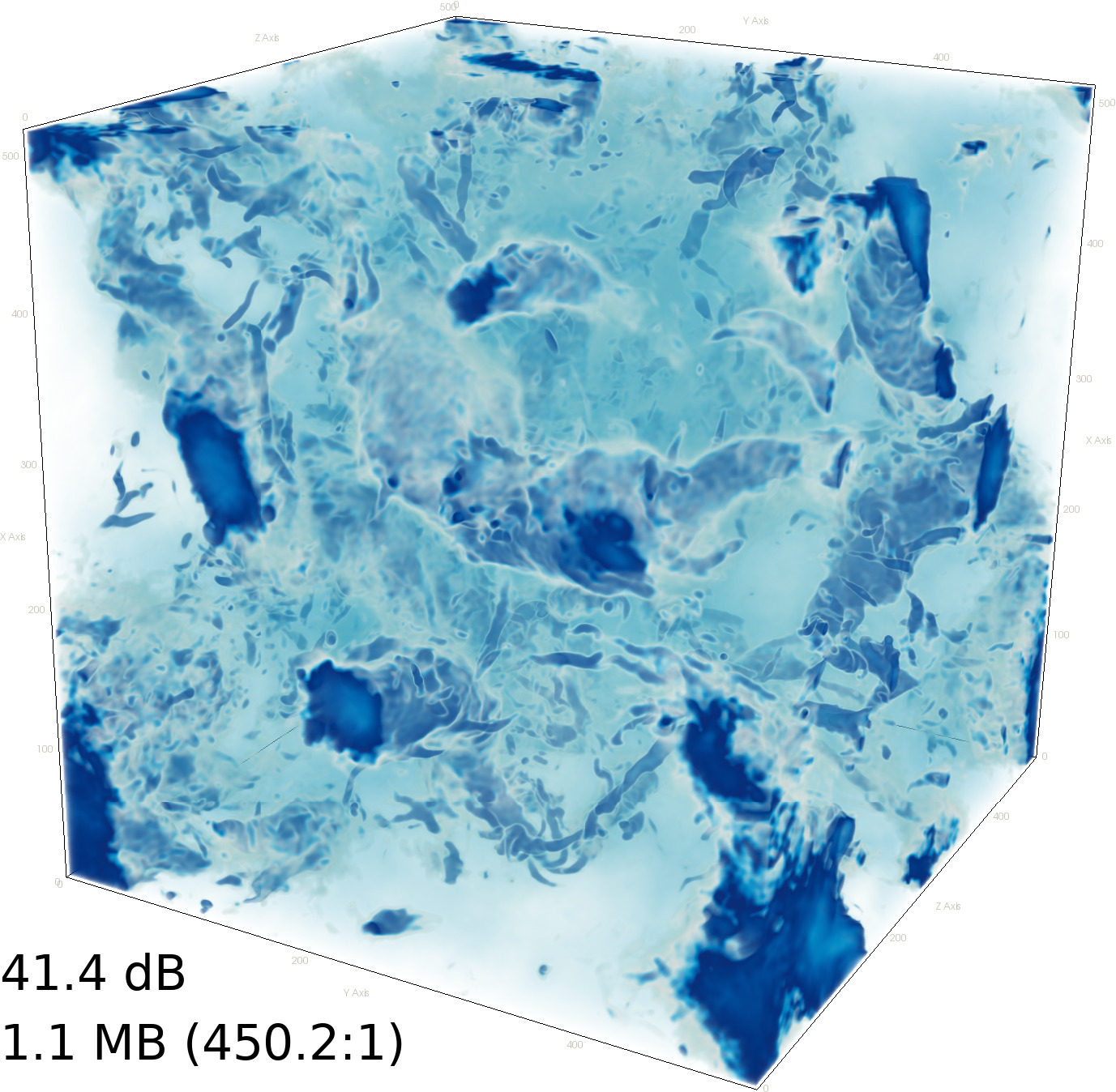}} \hfil
\subfigure{\includegraphics[width=0.47\columnwidth]{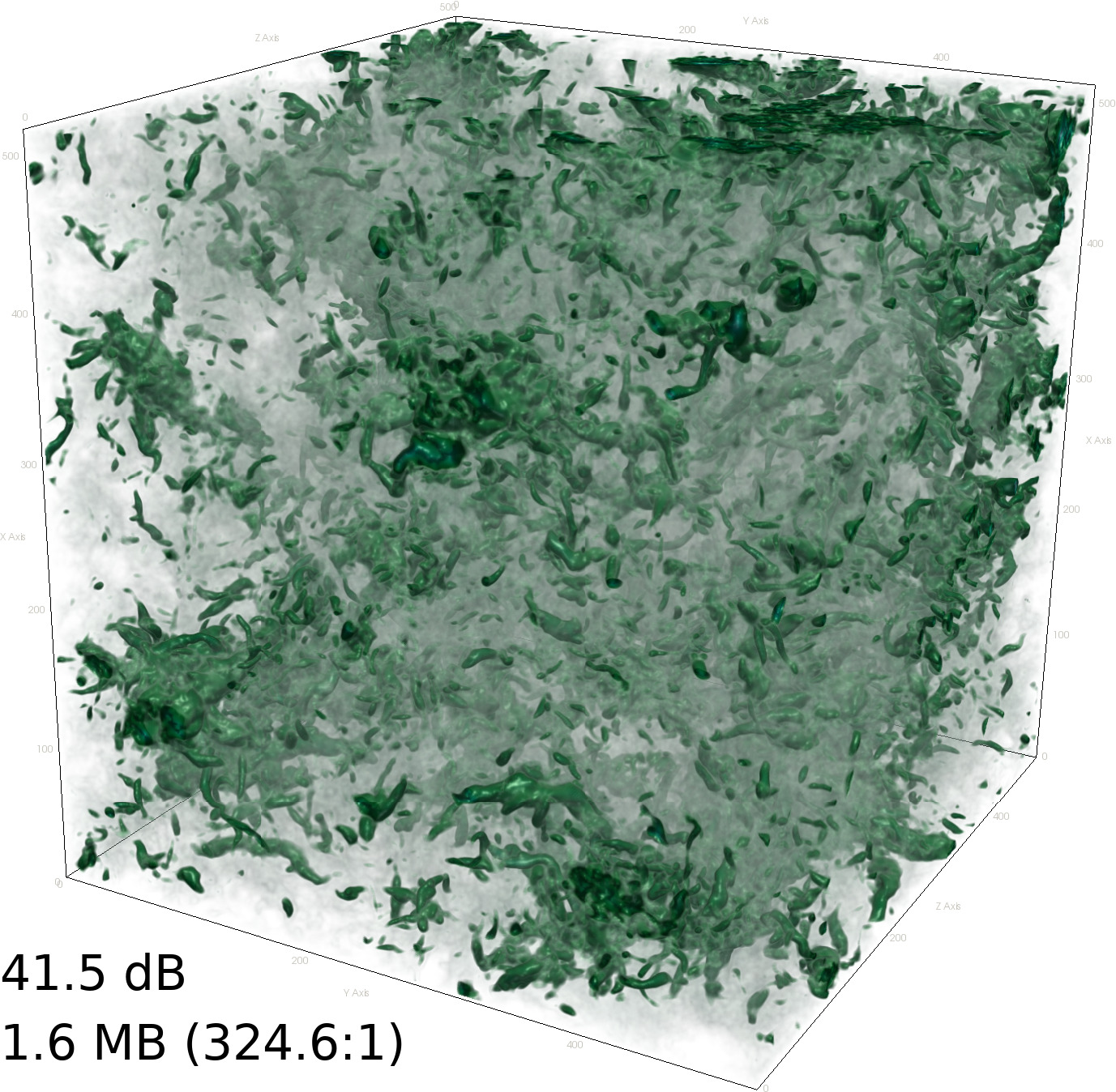}} \\
\caption{Four example volumes: the Foot and Engine CT scans (both 8-bit unsigned int), and the Isotropic-coarse and Mixing turbulence simulations (both 32-bit float). Rows from top to bottom: original, higher quality, and lower quality. All these compressed volumes take half or less the space needed with the other four methods tested at equivalent PSNR (except the Foot at higher quality, which performs similarly); see also Fig.~\ref{fig:tthresh_plots}.}
\label{fig:compression}
\end{figure*}

To support this claim we present several volume renderings before and after compression at two levels of quality in Fig.~\ref{fig:compression} as well as in the paper teaser (Fig~\ref{fig:teaser}).

We observe how the fixed number of quantization bits used by \textsc{trunc} entails a fixed error that often dominates that introduced by the rank truncation. This explains the flat PSNR curves for \textsc{trunc} over several data sets in Figs.~\ref{fig:tthresh_plots} and~\ref{fig:tthresh_plots2}. Interestingly, \textsc{sq} is sometimes not monotonic (see e.g. the results for the Channel volume). We attribute this to its set partitioning strategy~\cite{IKK:12}, whose resulting partitions can be highly sensitive to even small variations of the error tolerance specified.

\begin{figure*}[ht]
\centering
\stackunder[0pt]{\subfigure{\includegraphics[width=0.39\columnwidth]{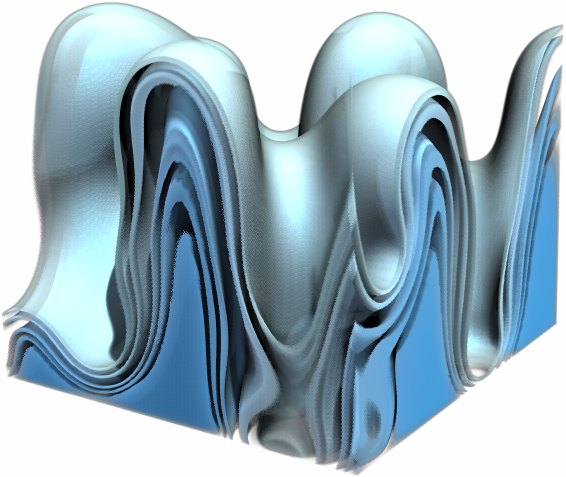}}}{\footnotesize Original $64^3$ brick} \hfil
\stackunder[0pt]{\subfigure{\includegraphics[width=0.39\columnwidth]{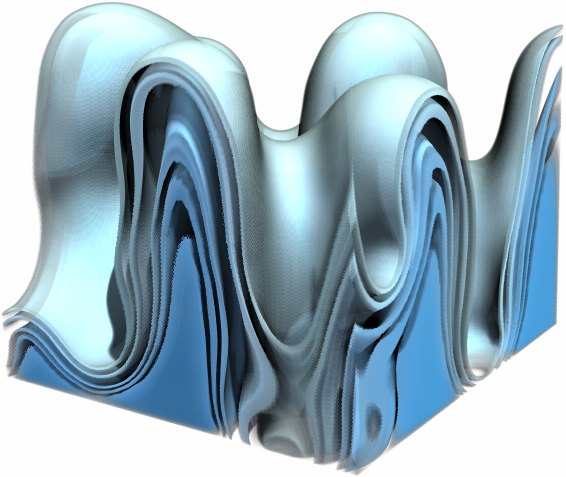}}}{\footnotesize \textsc{tthresh}, 47.1dB, 1,855:1} \hfil
\stackunder[0pt]{\subfigure{\includegraphics[width=0.39\columnwidth]{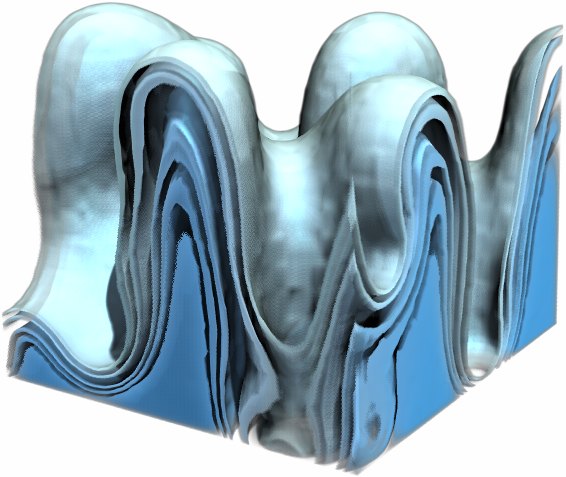}}}{\footnotesize \textsc{zfp}, 47.1dB, 107:1} \hfil
\stackunder[0pt]{\subfigure{\includegraphics[width=0.39\columnwidth]{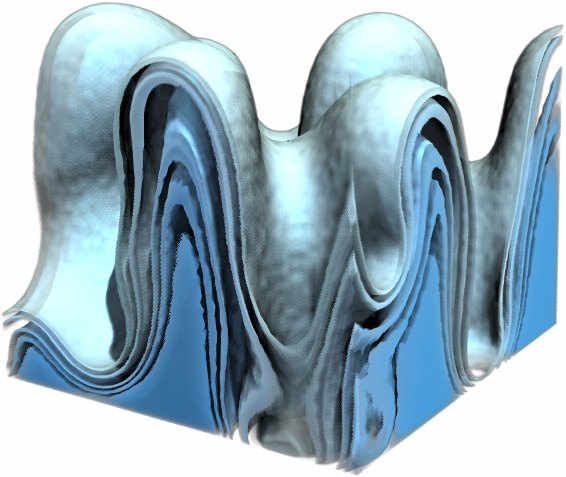}}}{\footnotesize \textsc{sz}, 46.0dB, 184:1} \hfil
\stackunder[0pt]{\subfigure{\includegraphics[width=0.39\columnwidth]{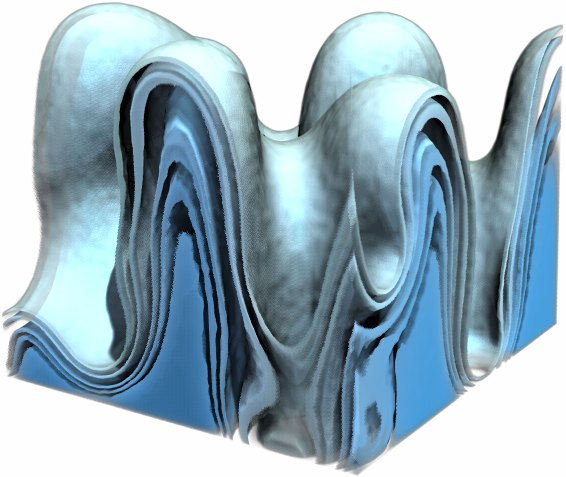}}}{\footnotesize \textsc{sq}, 46.1dB, 171:1} \\
\stackunder[0pt]{\subfigure{\includegraphics[width=0.39\columnwidth]{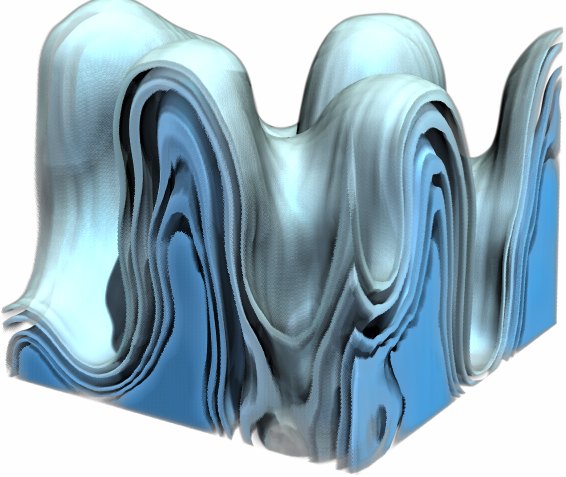}}}{\footnotesize \textsc{tthresh}, 30.3dB, 7,463:1} \hfil
\stackunder[0pt]{\subfigure{\includegraphics[width=0.39\columnwidth]{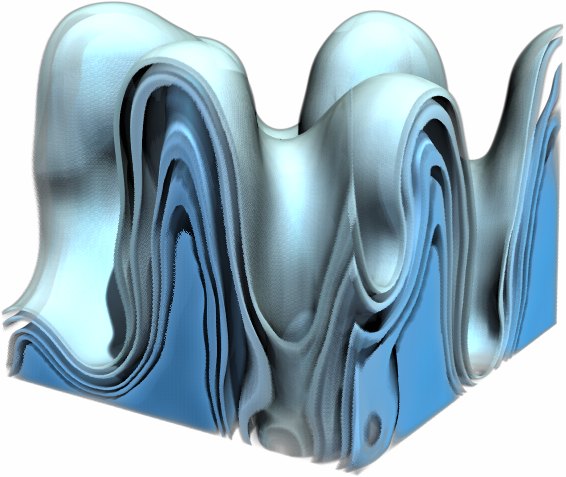}}}{\footnotesize \textsc{tthresh}, 38.6dB, 3,494:1} \hfil
\stackunder[0pt]{\subfigure{\includegraphics[width=0.39\columnwidth]{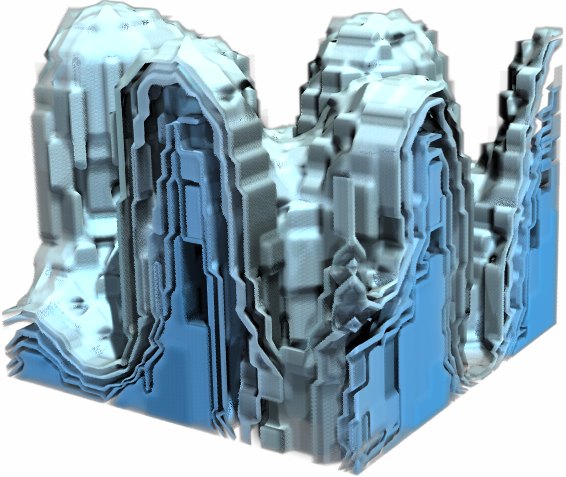}}}{\footnotesize \textsc{zfp}, 28.8dB, 160:1} \hfil
\stackunder[0pt]{\subfigure{\includegraphics[width=0.39\columnwidth]{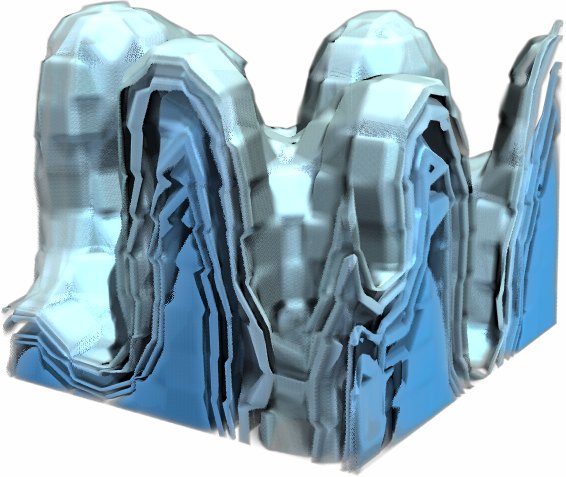}}}{\footnotesize \textsc{sz}, 30.3dB, 3,482:1}
\stackunder[0pt]{\subfigure{\includegraphics[width=0.39\columnwidth]{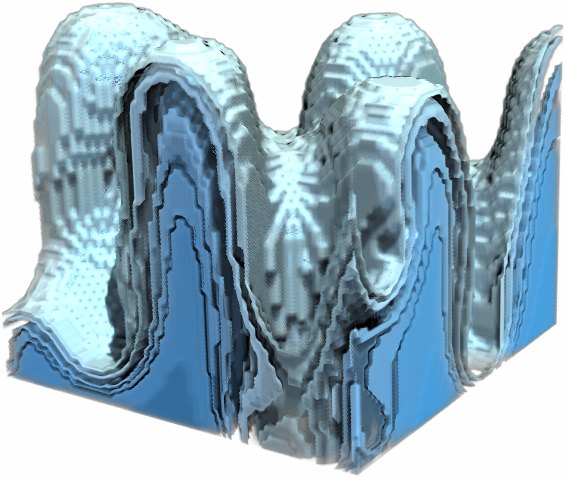}}}{\footnotesize \textsc{sq}, 30.1dB, 335:1}
\caption{The HOSVD produces custom data-dependent bases that make the proposed algorithm degrade visually very smoothly up to extreme compression rates. Depicted is a $64^3$ brick that was cut out from the center of the Density volume after compression with varying quality and algorithms (measurements correspond to the full volume). Unlike other compressors, \textsc{tthresh} avoids blocky artifacts; instead, it erodes and merges features at progressively coarser scales.}
\label{fig:features}
\end{figure*}

We also note that, at medium to high compression ratios, \textsc{tthresh} tends to preserve well the coarsest features and smoothen out or eliminate smaller details. See for example Fig.~\ref{fig:features} for a sequence of zoomed-in renderings under progressively heavier compression that make this phenomenon evident. It is only at exceedingly high ratios that block-like features start to appear, whereas other algorithms suffer from a much faster visual degradation. This multiscale feature-selective behavior is similar to that observed in truncation-based tensor compression~\cite{SIMAEZGGP:11}.

\begin{figure}[ht]
\centering
\subfigure[HOSVD factors]{\includegraphics[width=1\columnwidth]{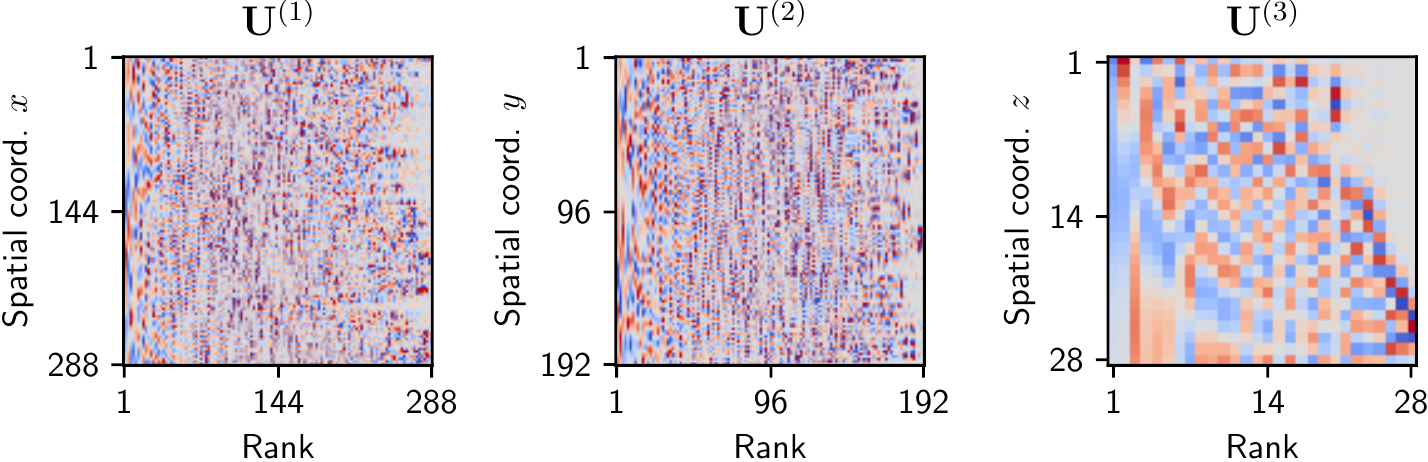}} \\
\subfigure[Their column-wise Fourier transform, in magnitude]{\includegraphics[width=1\columnwidth]{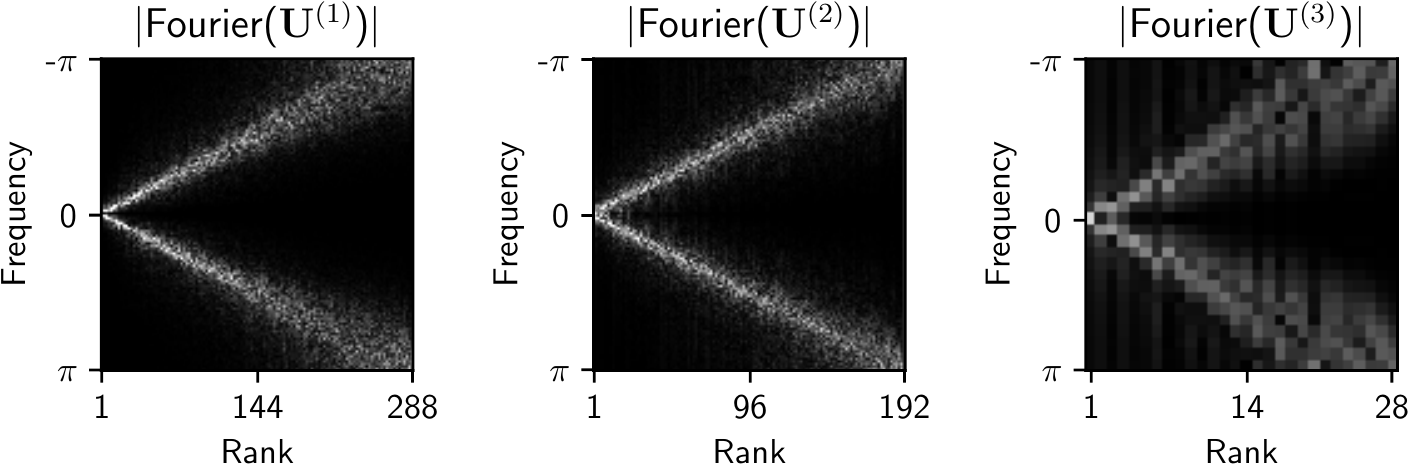}}
\caption{Factors obtained from the U data set along with their Fourier transform. HOSVD bases often resemble cosine wavefunctions and are rather sparse signals in the frequency domain.}
\label{fig:spectrum_matrices}
\end{figure}

It is further backed by empirical observation of the Fourier spectra of our compressed data sets. In Fig.~\ref{fig:spectrum_matrices} we show three example factor matrices for the U volume and their corresponding Fourier transforms along the spatial dimension (i.e. factor columns). Often, each vector in the HOSVD basis corresponds roughly to one frequency wavefunction that has been tuned to better match the specific input data set. Due to the \emph{hot corner} phenomenon, most insignificant HOSVD core coefficients are those that correspond to trailing factor columns (recall also Fig.~\ref{fig:slice_norms}) that, in light of Fig.~\ref{fig:spectrum_matrices}, contain mostly high frequencies. In short, we can expect our coder to act as a low-pass filter. This is consistent with what we showed in Fig.~\ref{fig:features}, namely smooth low-frequency artifacts that arise in data compressed with \textsc{tthresh}.

\begin{figure}[ht]
\centering
\subfigure[\textsc{tthresh}]{\includegraphics[width=0.45\columnwidth]{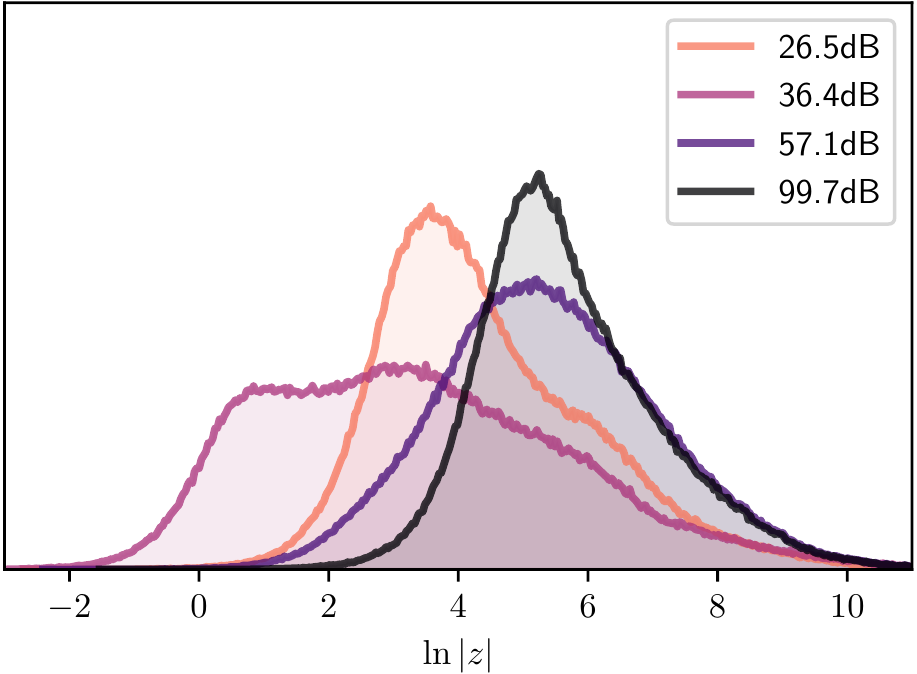}} \hfil
\subfigure[\textsc{zfp}]{\includegraphics[width=0.45\columnwidth]{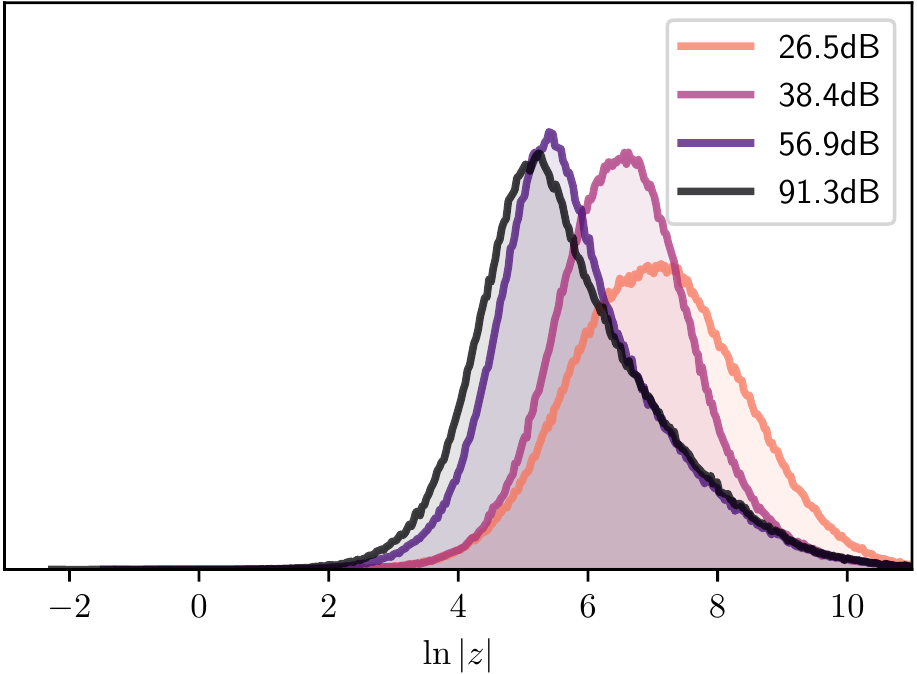}} \\
\subfigure[\textsc{sz}]{\includegraphics[width=0.45\columnwidth]{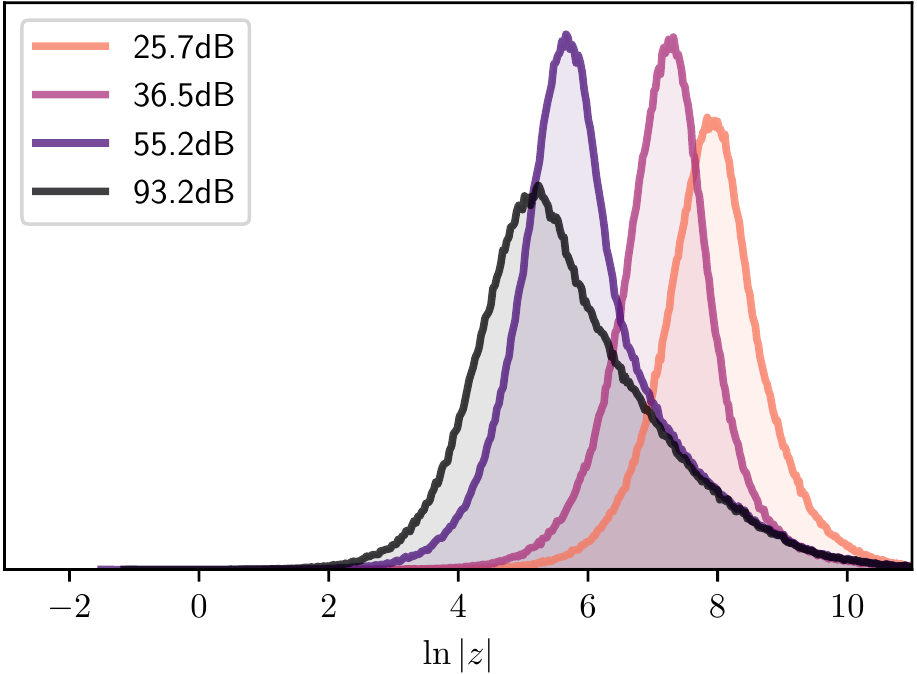}} \hfil
\subfigure[\textsc{sq}]{\includegraphics[width=0.45\columnwidth]{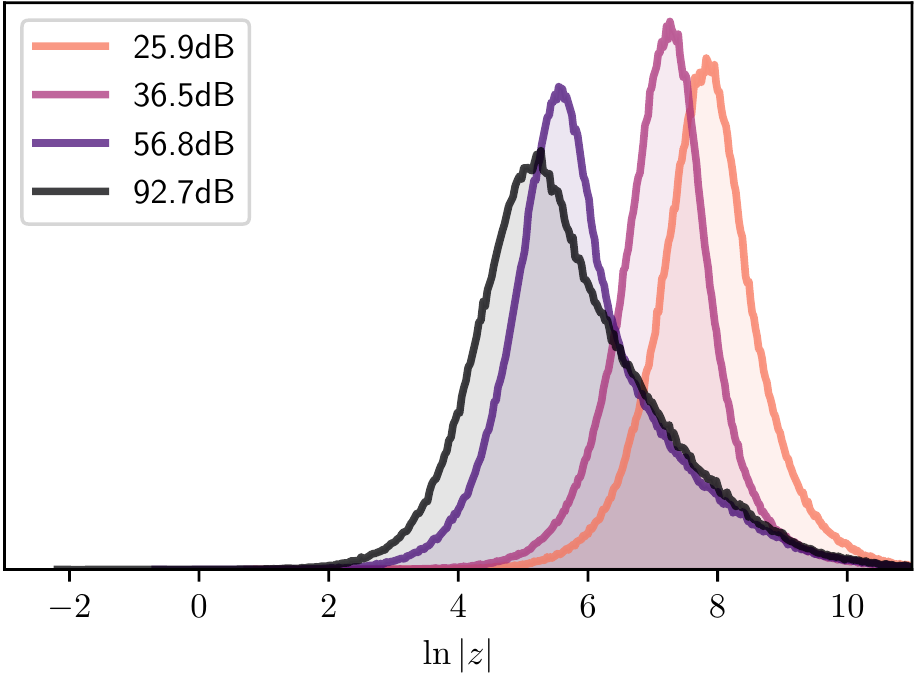}}
\caption{Logarithmic histograms of the Fourier transform magnitude at different compression quality levels of the U volume. \textsc{tthresh} shifts the spectrum towards lower frequencies, as opposed to the other three methods which tend to introduce higher frequencies instead.}
\label{fig:spectrum_histograms}
\end{figure}

To better illustrate this shift towards low frequencies, we depict in Fig.~\ref{fig:spectrum_histograms} the Fourier magnitude histograms obtained at different compression rates. Note that \textsc{sz}, \textsc{sq} and \textsc{zfp} behave in the opposite way as they rather shift the spectrum towards the high-frequency end.

\begin{figure}[ht]
\centering
\subfigure{\includegraphics[width=0.46\columnwidth]{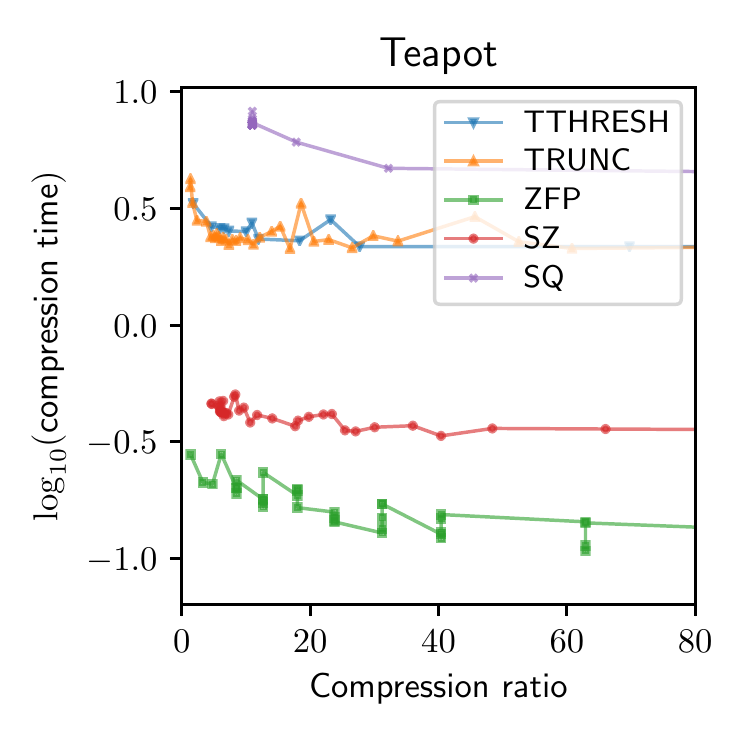}} \hfil
\subfigure{\includegraphics[width=0.46\columnwidth]{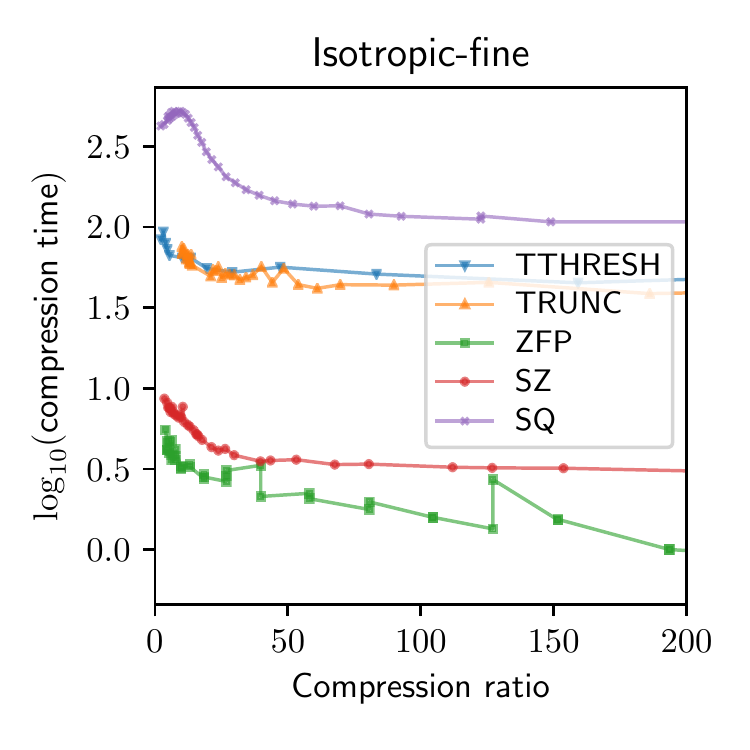}} \\
\subfigure{\includegraphics[width=0.46\columnwidth]{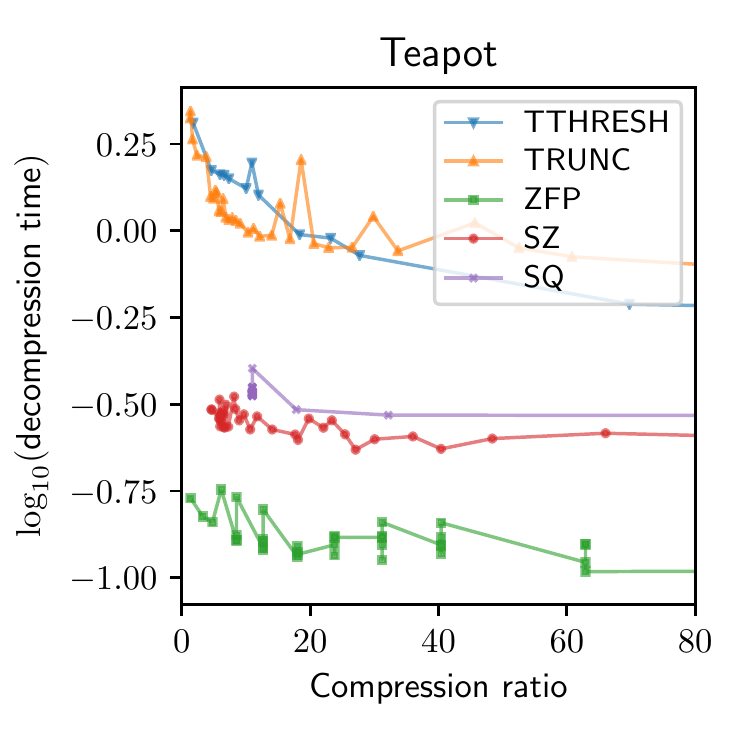}} \hfil
\subfigure{\includegraphics[width=0.46\columnwidth]{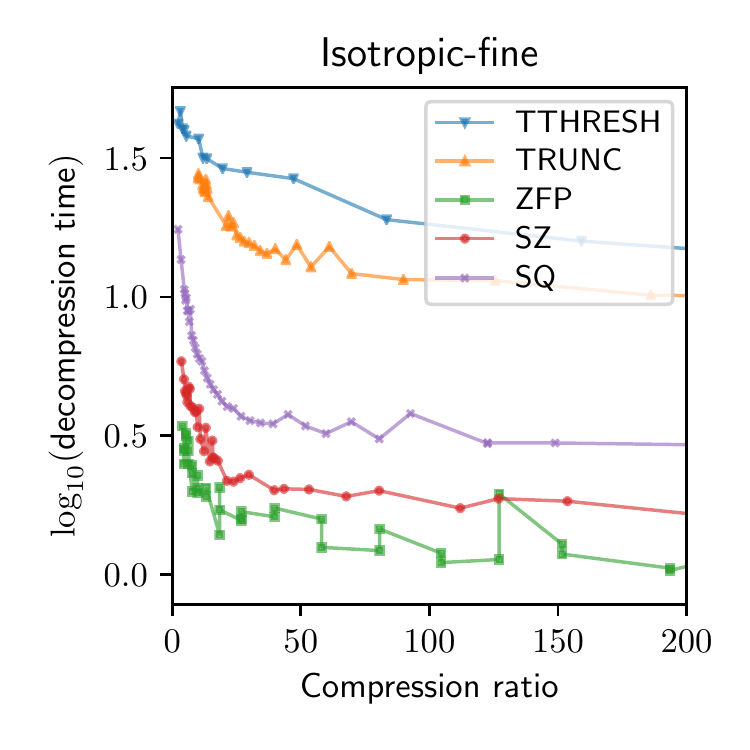}} \\
\caption{Compression (top row) and decompression (bottom row) times (in seconds) for two volumes and a range of different compression ratios.}
\label{fig:timings}
\end{figure}

\begin{figure}[ht]
\centering
\subfigure[Compression]{\includegraphics[width=0.45\columnwidth]{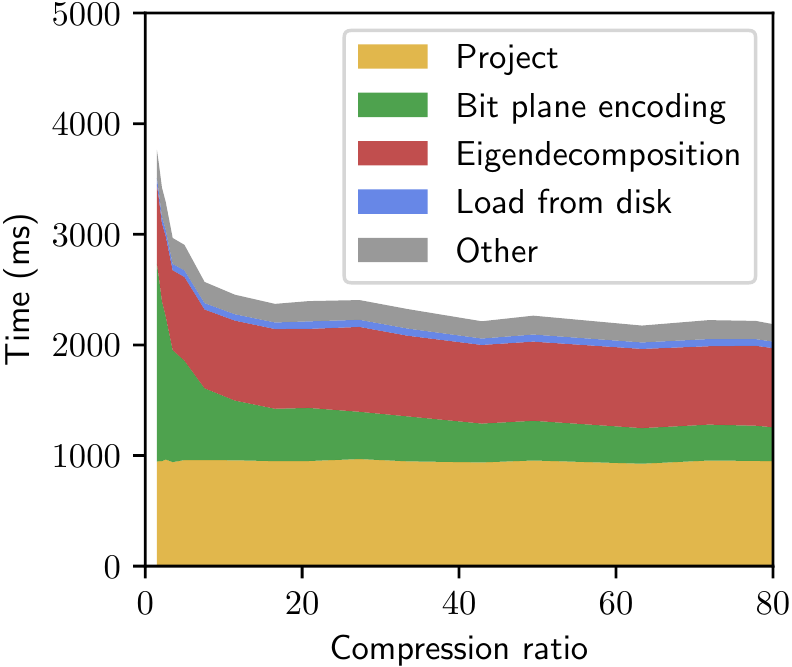}} \hfil
\subfigure[Decompression]{\includegraphics[width=0.45\columnwidth]{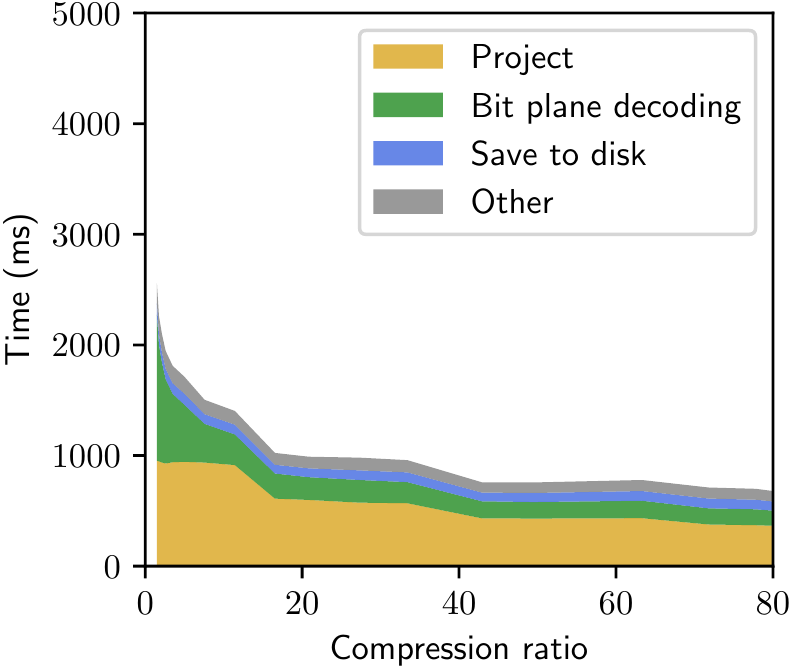}}
\caption{Compression/decompression times (Teapot): loading/saving the data set, eigenvalue decomposition, tensor projection with Tucker factors, and bit plane processing.}
\label{fig:profiling}
\end{figure}

Regarding computational speed, we plot in Fig.~\ref{fig:timings} the compression and decompression times for the smallest data set (the Teapot, 11.1MB) as well as for one of the 512MB ones (the Isotropic-fine). Our method is between 0.5 and 2 orders of magnitude slower than the fastest one, namely \textsc{zfp}, but compression is generally faster than \textsc{sq}. It is rather asymmetric as we expected from Sec.~\ref{sec:compression:decompression}: the average compression/decompression times for those measurements was 2.4s/1.0s (Teapot) and 61.5/25.8s (Isotropic-fine). To give more insight on the differences between compression and decompression costs, we have broken them down in Fig.~\ref{fig:profiling} (Teapot volume). Note also that the varying accuracy curves between all five compared algorithms make a fully fair comparison difficult. For consistency with Fig.~\ref{fig:tthresh_plots} we did the comparison in terms of time vs. compression ratio, but note that \textsc{tthresh} fares better in terms of quality in large parts of the error spectrum.

\begin{figure}[ht]
\centering
\subfigure[Original]{\includegraphics[width=0.24\columnwidth]{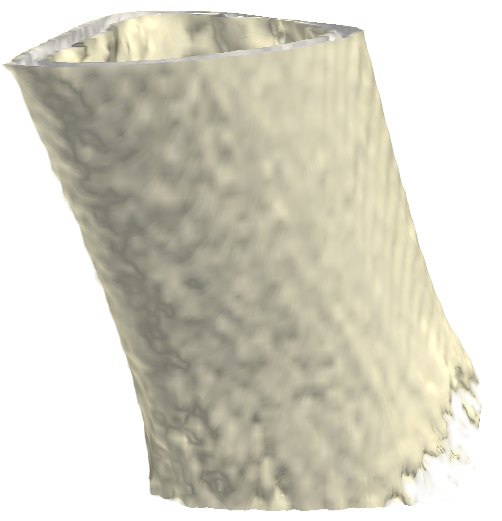}} \hfil
\subfigure[Downsampling]{\includegraphics[width=0.24\columnwidth]{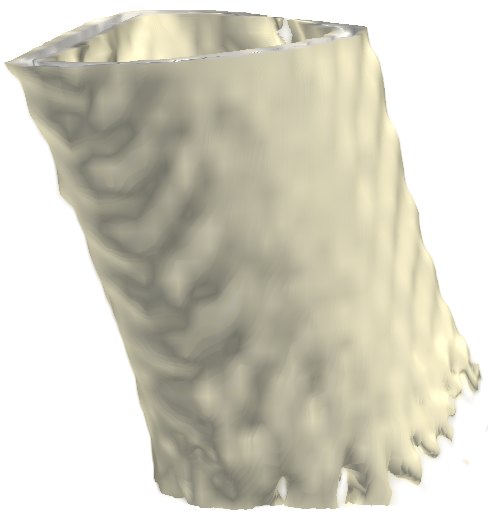}} \hfil
\subfigure[Box filter]{\includegraphics[width=0.24\columnwidth]{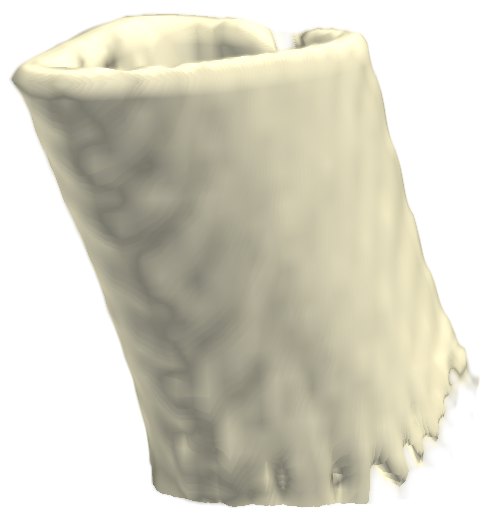}} \hfil
\subfigure[Lanczos-2]{\includegraphics[width=0.24\columnwidth]{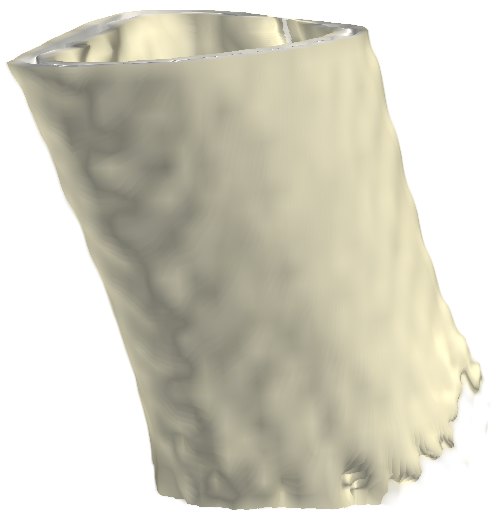}}
\caption{(a) a $60^3$ region of the Foot data set. (c-d): 2-fold decimated versions at 7.5:1 compression using the three different methods from Sec.~\ref{sec:compression:decompression}. Lanczos minimizes both the blocky aliasing along edges of pure downsampling and the erosion that the box filter incurs (top of the bone).}
\label{fig:downsampling}
\end{figure}

Our last experiment is reported in Fig.~\ref{fig:downsampling}, where we demonstrate visual results of decimation along the factor matrices (Sec.~\ref{sec:compression:decompression}) followed by decompression. Note the differences between the three methods implemented and the superiority of Lanczos' kernel for this task.

\section{Discussion}

We observe that the proposed algorithm achieves competitive accuracy at low to medium compression ratios and consistently outperforms other compressors at medium to high ratios; see the higher PSNR curves for our method in the highlighted regions of each plot from Fig.~\ref{fig:tthresh_plots}. The overtaking point at which \textsc{tthresh} surpasses the other algorithms (marked by the vertical dotted lines) typically produces renderings that are already close to visually indistinguishable to the original data set. This is especially true for higher bit depths. We believe our method is thus a good choice for applications with reasonable error tolerance (chiefly, visualization-related). In addition, we showed how our choice of global bases helps the method achieve a very smooth degradation rate. This is manifested both as a Fourier spectrum shift and as a visually parsimonious erosion of the smaller details and features. Since the transmission can be stopped at any arbitrary point within any bit plane, the range of possible final errors has a very fine granularity. Also, the error that arises from the core compression is upper-bounded by definition of the stopping criterion. Last, the compressed-domain filtering and resampling features are rather unique strengths of the tensor decomposition framework, only possible thanks to its multilinearity. Any separable filter and resampling can be applied with little cost by manipulating the factors column-wise before the final Tucker reconstruction. This is often much more challenging in other compression methods, especially brick-based and non-transform ones. Even though Lanczos antialiasing results are visually superior when lowering a data set's resolution, we believe the other decimation methods remain useful for other operations such as region/slice selection, projections, etc.

\subsection*{Limitations}

\textsc{tthresh}'s compression rates and smooth degradation come at the price of its monolithic approach to the transform core. This puts it in the slower end of the spectrum of volume compressors, especially compared to those designed for speed such as \textsc{zfp}. Random-access decompression is also relatively costly, as one must traverse the whole core in all cases. To improve compression/decompression speed one may resort to splitting the data set and using the proposed compressor on a brick-by-brick basis, in the spirit of tensor-compressed multiresolution rendering systems~\cite{SMP:13, BSP:18}.

\section{Conclusion} \label{sec:conclusions}

We have introduced a novel tensor decomposition based compression algorithm with an emphasis on storage/visualization applications whose foremost priority is data reduction at high compression ratios. Unlike previous HOSVD-driven approaches, this reduction is achieved by keeping all ranks followed by careful lossless compression of all bit planes up to a certain threshold. It is, to the best of our knowledge, the first tensor compressor (and specifically, HOSVD-based) that uses a bit-plane based strategy, also on the factor matrices. The main property we exploited was factor orthogonality, which ensures that all coefficients affect equally the final $l^2$ error and so allows us to sort the full core as a single block. Our algorithm possesses advantages that are inherent to multilinear transforms in general and tensor decompositions in particular, including support for linear manipulation of the data set in the compressed domain.

We developed \textsc{tthresh} focusing primarily on optimizing data reduction rates, and less so on general compression/decompression speed. We have realized that these speeds (especially compression) can be increased significantly at a relatively small accuracy cost in multiple ways, for example by moderating the eigensolver's number of iterations or by preemptively discarding some of the least important core slices. Also, we note that progressive decompression is compatible with the proposed coder: after all, we encode bit planes from more to less significant. To achieve progressiveness, coefficient signs should be encoded as soon as the coefficient becomes significant, e.g. using a negabinary base or deferred sign coding. In addition, factor columns should be encoded as soon as they are needed. These possibilities will be the subject of future investigation.



\acknowledgments{
This work was partially supported by the University of Zurich's Forschungskredit ``Candoc'', grant number FK-16-012, and partially performed under the auspices of the U.S. Department of Energy by Lawrence Livermore National Laboratory under Contract DE-AC52-07NA27344. The authors wish to thank Stephen Hamilton from Johns Hopkins University as well as the other institutions listed in Tab.~\ref{tab:datasets} for kindly providing the data sets we have used for testing.
}

\bibliographystyle{IEEEtran}
\bibliography{references}

\end{document}